# Adsorption of Phosgene Gas on Pristine and Noble Metal-Doped $B_{12}N_{12}$ Nanocages: Insights from Density Functional Theory

Shahariar Chowdhury[1,2], Mohammad Abdul Matin[2], Samiran Bhattacharjee[2*] and Ishtiaque M. Syed[1*]

[1]*Centre for Advanced Research in Sciences (CARS), University of Dhaka, Dhaka, Bangladesh*
[2]*Materials Physics Laboratory, Department of Physics, University of Dhaka, Dhaka, Bangladesh*

*Corresponding authors' e-mail: imsyed@du.ac.bd (IMS), s.bhattacharjee@du.ac.bd (SB)

## Abstract

This study examines phosgene ($COCl_2$) adsorption on pristine and noble metal-doped (Ag, Au, Pd, Pt) $B_{12}N_{12}$ nanocages using dispersion-corrected density functional theory [B3LYP-D3(BJ)]. Boron-site substitution narrows the HOMO–LUMO gap far more than nitrogen-site substitution (69–82% against 41–67%), so it was used throughout. All systems were fully optimised, and multiple starting geometries converged to two stable configurations per dopant, X–O and X–Cl, confirming these as the global minima. Pristine $B_{12}N_{12}$ binds phosgene weakly ($E_{ads}$ = −12.8 to −24.9 kJ mol$^{-1}$, that is −0.13 to −0.26 eV). Each of the four dopants binds it more strongly in its preferred orientation, the eight doped configurations spanning −19.5 kJ mol$^{-1}$ (Au–Cl) to −69.1 kJ mol$^{-1}$ (Pt–O); after counterpoise correction for basis-set superposition error the doped range is −14.0 to −60.7 kJ mol$^{-1}$. Platinum is the strongest adsorber both before and after that correction. Cohesive energies confirm that all doped cages remain thermodynamically stable, and platinum retains the most structural stability of the four. Natural population analysis shows that phosgene donates at most 0.33 e to the cage, and a QTAIM analysis of the wavefunctions classifies every cage···adsorbate bond critical point as closed-shell or intermediate, so the interaction is physisorption throughout. Vibrational analysis confirms every structure as a true minimum and gives the free energy of adsorption directly. The entropy cost of adsorption, 59–161 J mol$^{-1}$ K$^{-1}$, is decisive: with a quasi-harmonic treatment of the low-frequency modes, only Pt–O ($\Delta G$ = −8.8 kJ mol$^{-1}$) and Pd–O (−7.2 kJ mol$^{-1}$) adsorb phosgene spontaneously at 298 K, while pristine $B_{12}N_{12}$ does not bind it at all ($\Delta G$ = +23.8 and +27.9 kJ mol$^{-1}$). Pt–O desorbs above 69.6 °C, giving a practical regeneration window. Silver-doped complexes give the narrowest post-adsorption gaps and the highest electrophilicity, but cannot retain the analyte at room temperature. Platinum doping is therefore the effective strategy for phosgene detection with $B_{12}N_{12}$ nanocages.

## 1. Introduction

Gas sensors underpin environmental monitoring, industrial safety, and regulatory compliance. Conventional detection instruments are bulky, expensive, and poorly suited to real-time deployment. The demand is therefore for materials offering high sensitivity, low detection limits, rapid response and recovery, selectivity, and room-temperature operation. Recent work has surveyed a wide range of candidates: carbon nanotubes, graphene, metal and metal oxide nanoparticles, two-dimensional (2D) nanomaterials, and hybrid semiconductor nanostructures [1], [2]. Flexible, room-temperature wearable sensors built from 2D semiconductors are especially promising for portable detection. This class spans graphene derivatives, dichalcogenides, and MXenes. Their appeal lies in mechanical flexibility, high charge carrier mobility, and large active surface areas [3], [4], and even simple point defects can raise their gas response by orders of magnitude [5].

The discovery of carbon nanotubes by Iijima [6] accelerated nanotechnology and drove demand for nanoscale materials with tailored electronic and surface properties [7]. Nanocages, nanoclusters, and nanotubes now serve in optics, catalysis, sensing, adsorption, medicine, and electronic devices [8]. Inorganic analogues have followed. Boron nitride (BN) nanotubes and nanocages, and coaxial cubic AlN–BN composites [9], have all been synthesised and studied as functional nanostructures.

Fullerene-like $X_{12}Y_{12}$ clusters (X = Al, B, Ga; Y = As, N, P) are exceptionally stable and are known as “magic clusters” [10]. Toftlund and Jensen showed that $B_{12}N_{12}$ is more stable than the isoelectronic $C_{24}$ fullerene [11]. Such III–V clusters hold promise for light-emitting diodes and microelectronics. $B_{12}N_{12}$ and $Al_{12}N_{12}$ in particular combine physicochemical properties suited to a wide range of applications [12]. BN materials offer high thermal conductivity, large HOMO–LUMO gaps, thermal stability, low dielectric constants, and oxidation resistance [13]. Their chemistry is further shaped by the intrinsic charge separation between electron-deficient boron (Lewis acid) and electron-rich nitrogen (Lewis base) sites [14].

The geometry and stability of fullerene-like $(BN)_n$ nanoclusters have been studied extensively by theoretical methods; the stable structures combine tetragonal and hexagonal BN rings [8], [15], [16], [17]. Oku et al. synthesised $B_{12}N_{12}$ cages built from eight hexagonal and six tetragonal rings [18], and Fowler et al. identified $B_{12}N_{12}$ and $B_{16}N_{16}$ as particularly stable “magic” BN fullerenes [19].

The favourable electronic properties and structural tunability of $B_{12}N_{12}$ have since been exploited in DFT studies across a broad range of analytes. Gas-sensing work has examined pristine and doped $B_{12}N_{12}$ against halogenated species such as $Cl_2$ and $COCl_2$ [20] and toxic gases such as CO and NO [21]; a review covering 2011–2022 sets out the potential of pure and modified cages as toxic gas sensors [22], including for phosgene and thionyl chloride [23], and halogenated gases have been examined alongside interference studies [24]. Beyond gases, the cage adsorbs drug molecules such as fluorouracil, nitrosourea, pyrazinamide, metformin and favipiravir [12], [25], [26], stores and senses hydrogen when decorated with yttrium [15], and detects carbonyl sulfide and related hazardous species [8]. Transition metal decoration alters its nonlinear optical properties [27], Cu and Rh doping improves NO and $H_2S$ sensing [28], and salt adsorption has also been reported [29].

Phosgene ($COCl_2$) is a highly toxic, colourless gas. It has been in industrial use for over two centuries, since John Davy first synthesised it in 1812. Exposure causes severe injury, including pulmonary oedema after an asymptomatic latency period [30], [31]. Its odour threshold lies above hazardous concentrations, so smell is not a reliable warning [32]. Sensitive and reliable detection is therefore essential.

Phosgene sensor development has followed several routes. Carbon nitride and related nanostructures show enhanced sensitivity in DFT and experimental studies [23]. Noble metal-functionalised 2D materials have also been examined. Au-decorated $Ti_3C_2$ MXenes discriminate $COCl_2$ from CO, $H_2S$, $NH_3$, and $NO_2$ through physisorption-based transduction [3]. Graphene with ordered doping configurations [33], [34] and transition metal-doped silicon-based fullerenes [35] have been proposed as well. Experimental approaches include organic fluorescent probes for phosgene, mustard gas, and nerve agents [36], and bifunctional luminescent probes for dual-analyte detection of phosgene and chemical warfare agent simulants [37]. For BN systems, phosgene adsorption has been reported on pristine and copper-decorated $B_{12}N_{12}$ [38]. Most recently, Sousa and Varela Júnior screened the full first-row 3d series (TM = Sc–Zn) by doping, decoration, and encapsulation, and showed that several 3d metals sharply increase $COCl_2$ sensitivity [20]. Several early 3d metals, however, bind phosgene above 5 eV, far too strongly for practical sensor regeneration. Noble metal-doped $B_{12}N_{12}$ nanocages have not been examined systematically. These metals uniquely pair moderate binding strength with resistance to poisoning.

Noble metals (Ag, Au, Pd, Pt) are attractive dopants for gas sensing. Early transition metals bind adsorbates too strongly, which impedes desorption and sensor regeneration. Noble metals bind moderately, so adsorption–desorption cycles stay reversible. They also resist poisoning by the analyte and show strong affinity for oxygen-bearing molecules. Both traits are desirable for $COCl_2$ detection. As shown below, the four metals span a narrow adsorption energy range (~0.20–0.72 eV). This coherent group behaviour is absent from the wider 3d transition metal series.

Motivated by these considerations, the present study investigates the adsorption of phosgene on pristine and noble metal (Ag, Au, Pd, Pt)-doped $B_{12}N_{12}$ nanocages using density functional theory. Doping, rather than surface decoration, was adopted as the modification strategy. Substitution integrates the metal into the cage framework and gives the metal centre greater thermodynamic stability. Electronic and geometric properties were characterised before and after $COCl_2$ adsorption. The analysis covers interaction energies, HOMO–LUMO orbital structures, dipole moments, density of states, molecular electrostatic potentials, Bader's atoms-in-molecules theory, NCI-RDG analysis, and the vibrational thermochemistry of adsorption. Boron-site and nitrogen-site doping were compared directly to identify the better substitution site for phosgene sensing. Scheme 1 summarises the workflow: the pristine cage is doped at a boron site, phosgene is adsorbed through either its oxygen or its chlorine end, and each complex is assessed against binding, electronic, bonding and solvent descriptors.

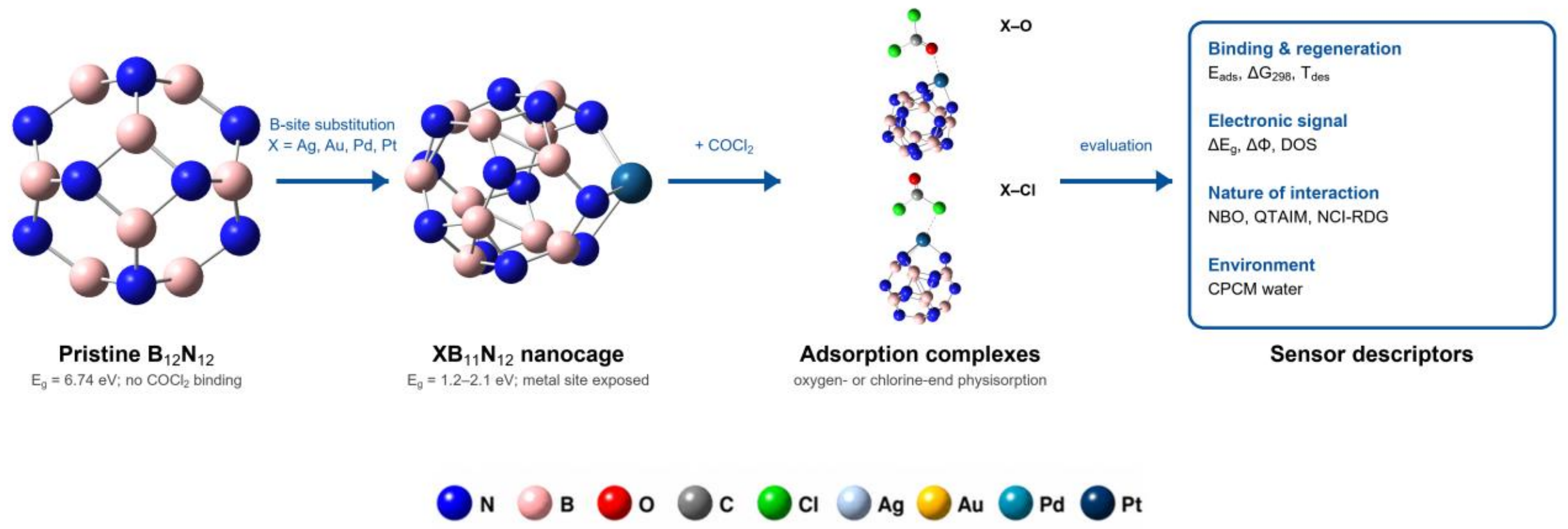

**Scheme 1.** Workflow of the present study. A boron atom of the pristine $B_{12}N_{12}$ cage is replaced by a noble metal X (Ag, Au, Pd, Pt); phosgene then adsorbs on the exposed metal site through its oxygen (X–O) or chlorine (X–Cl) end, and each complex is evaluated through its binding thermodynamics, desorption temperature and recovery time, its electronic signal, the nature of the cage···adsorbate interaction, and its behaviour in water.

## 2. Computational Details

### 2.1 Electronic structure calculations

All density functional theory (DFT) calculations used the Gaussian 16 program [39]. Geometries were optimised with the B3LYP hybrid exchange–correlation functional, augmented by Grimme's D3 dispersion correction with Becke–Johnson damping [B3LYP-D3(BJ)] [40], [41]. All runs used the 6-311++G(d,p) basis set for the main-group atoms (B, N, C, O, Cl); the metals (Ag, Au, Pd, Pt) were described with the LanL2DZ effective core potential (ECP) and its valence basis [42], and this composite is denoted GENECP throughout. Dipole moments, molecular electrostatic potentials (MEP), HOMO–LUMO gaps and density of states (DOS) were evaluated at the same B3LYP-D3(BJ)/GENECP level; α-spin frontier orbitals are used throughout for the open-shell Pd- and Pt-doped systems. Basis set superposition error was corrected with the counterpoise method [43]. Data extraction and wavefunction analysis used Multiwfn [44] and GaussSum [45].

Harmonic vibrational frequencies were computed for every optimised structure at the same level of theory. No imaginary frequencies were found, confirming all optimised structures as genuine minima on the potential-energy surface; the same frequencies supply the zero-point and thermal corrections used in Section 2.5.

For each doped nanocage, $COCl_2$ was placed in multiple starting orientations relative to the dopant centre: oxygen-facing, chlorine-facing, carbon-facing and side-on. Every starting geometry converged to either the X–O or the X–Cl configuration, so these two represent the global minima for phosgene adsorption on each doped system.

The level of theory was checked against the established properties of the pristine cage before the doped systems were studied. B3LYP-D3(BJ)/6-311++G(d,p) reproduces the $T_h$-symmetric $B_{12}N_{12}$ cage of six tetragonal and eight hexagonal rings, with B–N bond lengths of 1.436 Å (hexagon–hexagon) and 1.484 Å (hexagon–tetragon) and a HOMO–LUMO gap of 6.74 eV, in agreement with the geometries and gaps reported for this cluster at comparable hybrid-DFT levels [11], [14], [19], [46]. The weak, orientation-dependent physisorption of $COCl_2$ on the pristine cage found here is likewise consistent with earlier DFT studies of phosgene on $B_{12}N_{12}$ [20], [38]. B3LYP with the D3(BJ) correction and the LanL2DZ ECP for the noble metals is the combination used in most recent $B_{12}N_{12}$ sensing studies [20], [22], [28], which also allows the present results to be compared directly with that literature.

### 2.2 Stability and adsorption energies

The thermodynamic stability of each nanocage was assessed through its cohesive energy per atom,

$$E_{\text{coh}} = \frac{1}{P}\left(E_{\text{nanocage}} - x\,E_B - y\,E_N - z\,E_M\right) \quad (1)$$

where P = x + y + z is the total number of atoms, $E_{nanocage}$ is the total energy of the optimised nanocage, $E_B$, $E_N$ and $E_M$ are the energies of the isolated (spin-polarised) boron, nitrogen and metal atoms, and x, y and z are their stoichiometric counts. A negative $E_{coh}$ denotes a bound structure.

The adsorption energy of phosgene on a nanocage is

$$E_{\text{ads}} = E_{X-\text{BN}} - (E_{\text{BN}} + E_X) \quad (2)$$

where $E_{X–BN}$ is the total energy of the adsorbate–nanocage complex, $E_{BN}$ that of the optimised nanocage and $E_X$ that of the isolated $COCl_2$ molecule. Adsorption energies were also corrected for basis set superposition error (BSSE) with the counterpoise method [42], and zero-point energy (ZPE) corrections were taken from the vibrational analysis.

### 2.3 Frontier orbitals and global reactivity descriptors

The HOMO–LUMO energy gap and the relative electronic sensitivity of a nanocage to adsorption are defined as

$$E_g = E_{\text{LUMO}} - E_{\text{HOMO}} \quad (3)$$

$$\%\Delta E_g = \frac{E_g(\text{complex}) - E_g(\text{nanocage})}{E_g(\text{nanocage})} \times 100\% \quad (4)$$

Global reactivity descriptors were obtained within conceptual DFT using the Koopmans' theorem approximations for the ionisation potential and electron affinity,

$$\text{IP} = -E_{\text{HOMO}} \quad (5)$$

$$\text{EA} = -E_{\text{LUMO}} \quad (6)$$

from which the electronic chemical potential μ, electronegativity χ, chemical hardness η, global softness S and electrophilicity index ω follow as

$$\mu = \frac{E_{\text{HOMO}} + E_{\text{LUMO}}}{2} = -\frac{\text{IP} + \text{EA}}{2} \quad (7)$$

$$\chi = -\mu = \frac{\text{IP} + \text{EA}}{2} \quad (8)$$

$$\eta = \frac{E_{\text{LUMO}} - E_{\text{HOMO}}}{2} = \frac{\text{IP} - \text{EA}}{2} \quad (9)$$

$$S = \frac{1}{2\eta} \tag{10}$$

$$\omega = \frac{\mu^2}{2\eta} = \frac{\chi^2}{2\eta} \tag{11}$$

**2.4 Fermi level, work function and solvation**

The Fermi-level energy is approximated by the midpoint of the frontier orbitals, which coincides with the chemical potential,

$$E_F = \frac{E_{\mathrm{HOMO}} + E_{\mathrm{LUMO}}}{2} = \mu \tag{12}$$

and the work function is the energy required to remove an electron from the Fermi level to the vacuum,

$$\Phi = V_{\mathrm{el}}(+\infty) - E_F \tag{13}$$

where $V_{el}(+\infty)$ is the vacuum electrostatic potential. Setting $V_{el}(+\infty) = 0$ gives $\Phi = -E_F$.

Solvent effects were evaluated with the conductor-like polarizable continuum model (CPCM) in water, as single-point calculations on the gas-phase geometries at the same level of theory. The solvation energy of each species is

$$E_{\mathrm{sol}} = E_{\mathrm{water}} - E_{\mathrm{gas}} \tag{14}$$

where $E_{water}$ and $E_{gas}$ are the total energies in solution and in the gas phase, respectively.

**2.5 Thermochemistry of adsorption**

Rigid-rotor harmonic-oscillator (RRHO) statistical thermodynamics at 298.15 K and 1 atm, built on the computed frequencies, gives the enthalpy, entropy and Gibbs free energy of adsorption as

$$\begin{aligned} \Delta H_{\mathrm{ads}} &= H(\mathrm{complex}) - H(\mathrm{cage}) - H(\mathrm{COCl_2}) \\ \Delta S_{\mathrm{ads}} &= S(\mathrm{complex}) - S(\mathrm{cage}) - S(\mathrm{COCl_2}) \\ \Delta G_{\mathrm{ads}} &= \Delta H_{\mathrm{ads}} - T\Delta S_{\mathrm{ads}} \end{aligned} \tag{15}$$

Adsorption converts the three translational and three rotational degrees of freedom of free phosgene into six very soft vibrations, so the complexes carry modes as low as 3 $cm^{-1}$. The harmonic vibrational entropy of a mode diverges as its frequency approaches zero and is therefore unreliable for these modes. Entropies are consequently also reported in Grimme's quasi-harmonic approximation, in which each mode is interpolated between a harmonic oscillator and a free rotor with the weighting $w(\omega) = [1 + (\omega_0/\omega)^4]^{-1}$ and $\omega_0 = 100$ $cm^{-1}$. The resulting $\Delta G_{qh}$ is used for all thermodynamic analysis below. The desorption temperature $T_{des}$ is the temperature at which $\Delta G_{ads}$ changes sign, obtained by recomputing the full partition functions as a function of temperature. Equilibrium constants follow as $K = \exp(-\Delta G_{qh}/RT)$ and Langmuir coverages as $\theta = Kp/(1 + Kp)$.

## 3. Results and Discussion

### 3.1 Pristine $B_{12}N_{12}$ and Initial Adsorption Geometries of $COCl_2$

Figure 1(a) shows the optimised pristine $B_{12}N_{12}$ nanocage at the B3LYP-D3(BJ)/GENECP level, and Figure 1(b) its molecular electrostatic potential. The cluster contains six tetragonal (4-membered) and eight hexagonal (6-membered) BN rings. Its stability arises from ionic bonding between positively charged B and negatively charged N atoms. B–N bond lengths depend on ring adjacency: $b_{66}$ = 1.436 Å for the twelve bonds shared by two hexagonal rings, and $b_{64}$ = 1.484 Å for the twenty-four bonds shared by a hexagonal and a tetragonal ring. Both agree with earlier theoretical and experimental work on this cage [11], [14], [18], [19], [46]. Natural bond orbital (NBO) analysis gives a charge of +1.155 e on every boron atom and −1.155 e on every nitrogen atom; the Mulliken partition gives the much smaller values of +0.179 e and −0.179 e, so the two schemes agree on the direction of the polarisation but not on its magnitude. The cohesive energy is −5.997 eV per atom. The preference for n = 12 among $(BN)_n$ cages is well established, and $B_{12}N_{12}$ was therefore adopted as the model nanocage throughout. The optimised free phosgene molecule and its MEP are shown in Figures 1(c) and 1(d): the carbonyl oxygen carries the electron-rich (red) region and the chlorine atoms the electron-poor (blue) belts, which is what steers its approach to the cage.

Phosgene ($COCl_2$) was approached to the pristine cage from multiple orientations, and two stable configurations emerged. In B–O the carbonyl oxygen faces the nearest boron atom (B···O = 2.34 Å), Figure 1(e), with the MEP of the complex in Figure 1(f). In B–Cl a chlorine atom points toward the cage surface, the shortest contact being N···Cl = 3.05 Å, Figures 1(g) and 1(h); see also Table 5. Cohesive energies, frontier-orbital energies, gaps and adsorption energies at the B3LYP-D3(BJ)/GENECP level are collected in Table 1, which also lists the corresponding properties of the doped nanocages discussed in Section 3.2.

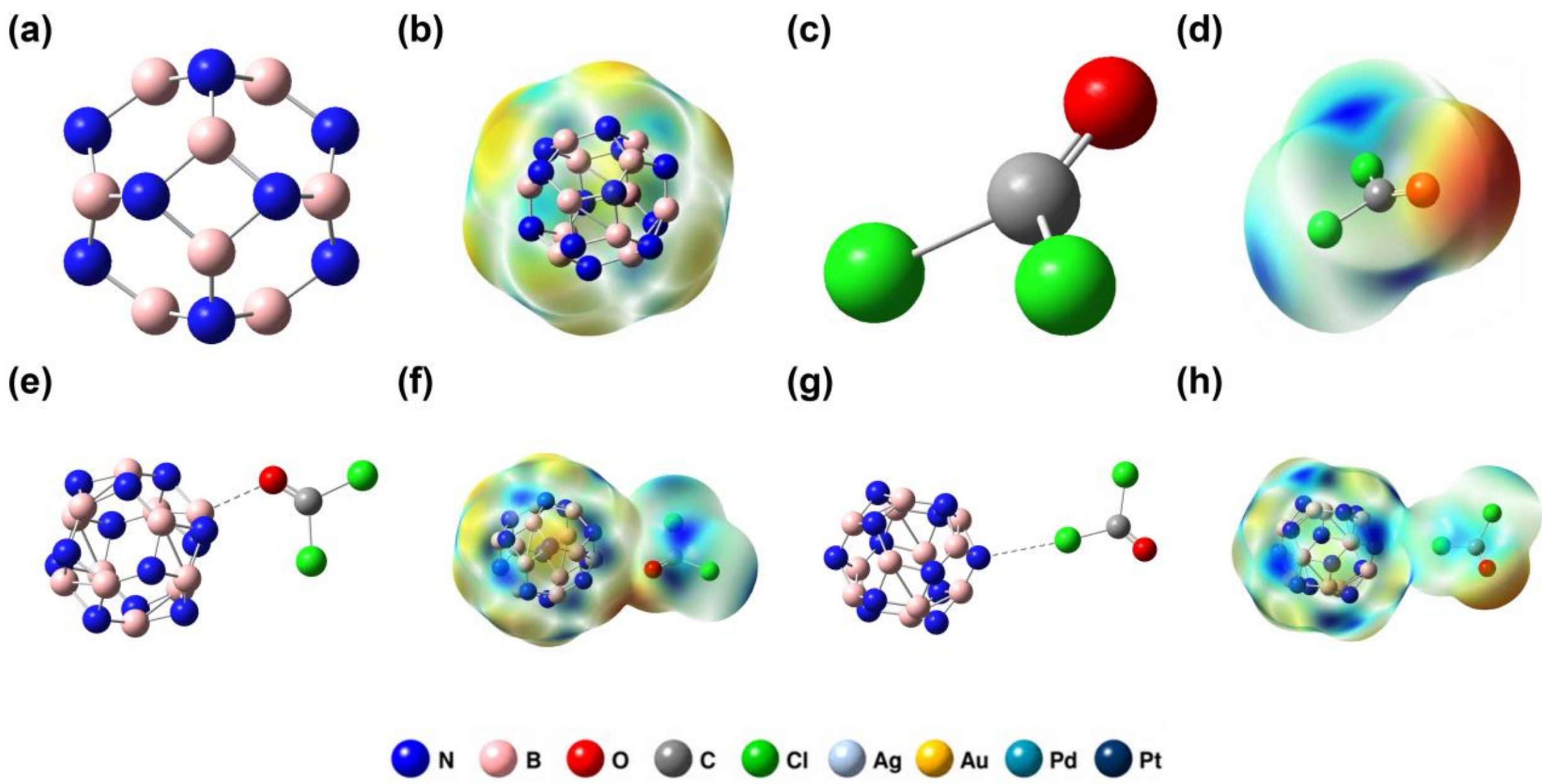

**Figure 1.** Representations of the optimised geometry of $B_{12}N_{12}$ (a), MEP of $B_{12}N_{12}$ (b), optimised structure of $COCl_2$ (c), MEP of $COCl_2$ (d), optimised structure of the $B_{12}N_{12}$–$COCl_2$ B–O complex (e), MEP of the B–O complex (f), optimised structure of the $B_{12}N_{12}$–$COCl_2$ B–Cl complex (g) and MEP of the B–Cl complex (h).

**Table 1.** Cohesive energy ($E_{coh}$), shortest cage···adsorbate contact, HOMO and LUMO energies ($E_H$, $E_L$), HOMO–LUMO gap ($E_g$), electronic sensitivity (%$\Delta E_g$), Fermi energy ($E_F$), adsorption energy ($E_{ads}$) and dipole moment (μ) of pristine $B_{12}N_{12}$, its two $COCl_2$ complexes and the boron- and nitrogen-site doped nanocages in the gas phase. %$\Delta E_g$ is relative to pristine $B_{12}N_{12}$.

| System | $E_{coh}$ (eV/atom) | Contact (Å) | $E_H$ (eV) | $E_L$ (eV) | $E_g$ (eV) | %$\Delta E_g$ | $E_F$ (eV) | $E_{ads}$ (kJ mol$^{-1}$) | μ (D) |
|---|---|---|---|---|---|---|---|---|---|
| ***Pristine nanocage and its $COCl_2$ complexes*** | | | | | | | | | |
| $B_{12}N_{12}$ | −5.997 | -- | −7.97 | −1.23 | 6.74 | -- | −4.60 | -- | 0.00 |
| B–O | −5.671 | 2.34 (B···O) | −7.74 | −2.50 | 5.24 | −22.24 | −5.12 | −24.90 | 2.56 |
| B–Cl | −5.667 | 3.05 (N···Cl) | −8.06 | −1.98 | 6.08 | −9.85 | −5.02 | −12.82 | 1.65 |
| ***Boron-site doping ($XB_{11}N_{12}$)*** | | | | | | | | | |
| $AgB_{11}N_{12}$ | −5.587 | -- | −6.73 | −5.55 | 1.19 | −82.43 | −6.14 | -- | 2.94 |
| $AuB_{11}N_{12}$ | −5.593 | -- | −6.76 | −5.30 | 1.46 | −78.38 | −6.03 | -- | 1.87 |
| $PdB_{11}N_{12}$ | −5.674 | -- | −6.64 | −4.58 | 2.06 | −69.45 | −5.61 | -- | 2.24 |
| $PtB_{11}N_{12}$ | −5.726 | -- | −6.33 | −4.23 | 2.10 | −68.85 | −5.28 | -- | 1.52 |
| ***Nitrogen-site doping ($XN_{11}B_{12}$)*** | | | | | | | | | |
| $AgN_{11}B_{12}$ | -- | -- | −6.18 | −2.25 | 3.94 | −41.34 | −4.22 | -- | -- |
| $AuN_{11}B_{12}$ | -- | -- | −5.18 | −2.72 | 2.46 | −63.33 | −3.95 | -- | -- |
| $PdN_{11}B_{12}$ | -- | -- | −4.79 | −2.41 | 2.38 | −64.53 | −3.60 | -- | -- |
| $PtN_{11}B_{12}$ | -- | -- | −4.78 | −2.54 | 2.23 | −66.73 | −3.66 | -- | -- |

Cohesive energies and dipole moments were not computed for the nitrogen-site series, which was screened on frontier-orbital energies alone.

The adsorption energies are −24.90 kJ mol$^{-1}$ (B–O) and −12.82 kJ mol$^{-1}$ (B–Cl), that is −0.258 and −0.133 eV; counterpoise correction reduces them to −17.55 and −8.57 kJ mol$^{-1}$. Both indicate weak van der Waals interaction with the pristine cage, and the separations of 2.3–3.1 Å support this. B–O is the more stable of the two, but neither is strong enough for practical sensing. Adsorption nevertheless narrows the gap of the pristine cage, from 6.74 eV to 5.24 eV in B–O (−22.24%) and to 6.08 eV in B–Cl (−9.85%). This is the opposite of the behaviour found for the doped cages below, where adsorption widens the gap in every case.

### 3.2 Comparative Analysis of Boron-Site and Nitrogen-Site Doping

Two substitution strategies were tested for each dopant X = Ag, Au, Pd, Pt: replacement of a boron atom (X@B, giving $XB_{11}N_{12}$) and replacement of a nitrogen atom (X@N, giving $XN_{11}B_{12}$). Optimised geometries of the boron-site series are shown in the top row of Figure 2 ($a_1$–$d_1$) and those of the nitrogen-site series in Figure S1 of the Supplementary Information. The corresponding electronic structure parameters are summarised in Table 1; cohesive energies and dipole moments were not computed for the nitrogen-site series, which was screened on frontier-orbital energies alone.

Pristine $B_{12}N_{12}$ has a HOMO–LUMO gap of 6.74 eV, typical of a wide-gap semiconductor and in line with previous hybrid-DFT values for this cluster [19], [22], [46]. Boron-site substitution ($XB_{11}N_{12}$) narrows this gap sharply for every dopant. $E_g$ falls to 1.19 eV for $AgB_{11}N_{12}$, an 82.43% reduction, and to 1.46 eV for $AuB_{11}N_{12}$ (78.38%). $PdB_{11}N_{12}$ and $PtB_{11}N_{12}$ give reductions of 69.45% and 68.85%, to 2.06 and 2.10 eV. Nitrogen-site substitution is far weaker. $AgN_{11}B_{12}$ falls only 41.34% ($E_g$ = 3.94 eV), and Au, Pd, and Pt at nitrogen sites give intermediate reductions of 63.33%, 64.53%, and 66.73%.

Boron sites perform better because boron is electron-deficient. Replacing it with a metal carrying partially filled d-orbitals introduces low-lying mid-gap states that bridge the HOMO–LUMO gap. The LUMO energies show this directly: for silver the LUMO drops from −2.25 eV ($AgN_{11}B_{12}$) to −5.55 eV ($AgB_{11}N_{12}$), reflecting the new low-lying unoccupied states. Boron-site doping also shifts the Fermi level to more negative values, −5.28 to −6.14 eV against −3.60 to −4.22 eV for nitrogen sites, and electron-accepting capability rises accordingly. Accordingly, only the boron-site doped series ($XB_{11}N_{12}$) was carried forward for $COCl_2$ adsorption studies.

The dopant distorts the $B_{12}N_{12}$ framework noticeably (Figure 2, $a_1$–$d_1$): the metal protrudes out of the cage surface and the surrounding rings deform to relieve torsional strain. All cohesive energies in Table 1 are negative, so every doped cage is thermodynamically stable. The pristine cage is the most stable ($E_{coh}$ = −5.997 eV per atom), as expected. $PtB_{11}N_{12}$ (−5.726 eV per atom) departs least from that value, indicating the smallest structural perturbation, followed by $PdB_{11}N_{12}$ (−5.674), $AuB_{11}N_{12}$ (−5.593) and $AgB_{11}N_{12}$ (−5.587). Metal doping narrows the HOMO–LUMO gap to 1.19–1.46 eV (Ag, Au) and 2.06–2.10 eV (Pd, Pt), corresponding to reductions of 78–82% and about 69% relative to the pristine cage.

### 3.3 $COCl_2$ Adsorption on Doped Nanocages

Each doped system gives two adsorption configurations. In X–O the carbonyl oxygen of $COCl_2$ coordinates to the dopant atom; in X–Cl a chlorine atom faces the dopant (X = Ag, Au, Pd, Pt). As described in the Computational Details, multiple initial orientations were tested and all converged to these two, confirming them as the global minima. Figure 2 shows the bare doped cages ($a_1$–$d_1$), the optimised oxygen-side complexes ($a_2$–$d_2$) and their MEP maps ($a_3$–$d_3$); the corresponding chlorine-side complexes and MEP maps are given in Figure S2.

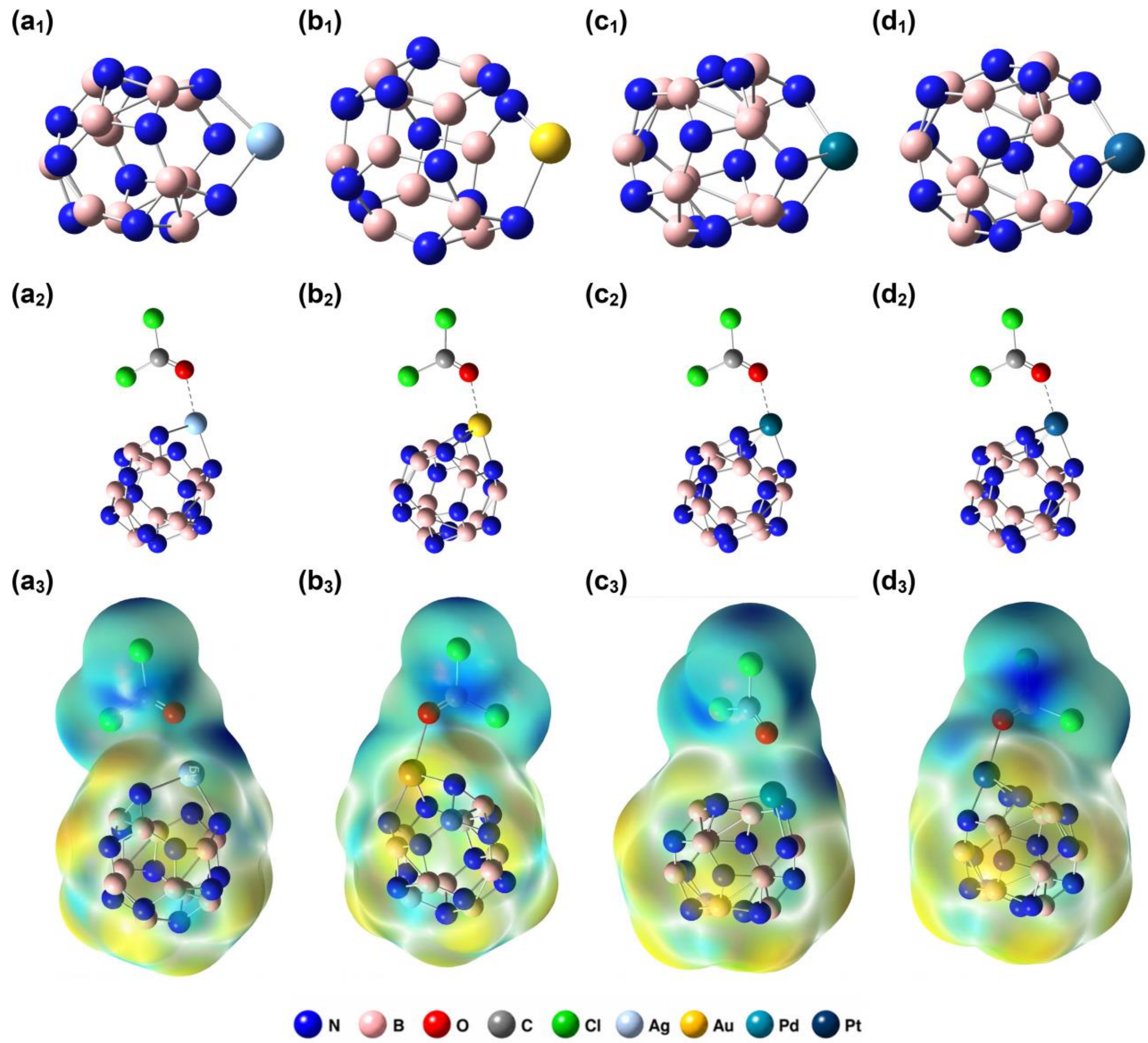

**Figure 2.** Representations of the optimised geometry of $AgB_{11}N_{12}$ ($a_1$), $AuB_{11}N_{12}$ ($b_1$), $PdB_{11}N_{12}$ ($c_1$) and $PtB_{11}N_{12}$ ($d_1$); optimised geometry of $AgB_{11}N_{12}$ ($a_2$), $AuB_{11}N_{12}$ ($b_2$), $PdB_{11}N_{12}$ ($c_2$) and $PtB_{11}N_{12}$ ($d_2$) with phosgene gas ($COCl_2$) adsorbed near the oxygen side; and MEP of $AgB_{11}N_{12}$ ($a_3$), $AuB_{11}N_{12}$ ($b_3$), $PdB_{11}N_{12}$ ($c_3$) and $PtB_{11}N_{12}$ ($d_3$) with phosgene gas ($COCl_2$) adsorbed near the oxygen side.

Table 2 collects the cohesive energies, contact distances, frontier-orbital energies, gaps, adsorption energies, dispersion contributions and dipole moments of free phosgene and the eight doped adsorption complexes in the gas phase; the corresponding data for the bare cages and the two pristine complexes are in Table 1. $COCl_2$ binding shifts the HOMO–LUMO gap of each doped cage substantially. Relative to the corresponding bare doped cage, the oxygen-facing (X–O) configurations change $E_g$ by +15.25% (Ag–O), +23.36% (Au–O), +21.70% (Pd–O), and +18.42% (Pt–O). The chlorine-facing set gives +11.83% (Ag–Cl), +2.03% (Au–Cl), +16.11% (Pd–Cl), and +53.87% (Pt–Cl). Conductivity in a semiconductor falls exponentially with the HOMO–LUMO gap [47]:

$$\sigma = \sigma_0 \exp\left(-\frac{E_g}{2kT}\right) \quad (16)$$

The increase in $E_g$ upon $COCl_2$ adsorption on the doped nanocages therefore results in a measurable conductivity change that can serve as the sensing signal.

**Table 2.** Cohesive energy ($E_{coh}$), interaction bond length, HOMO and LUMO energies, HOMO–LUMO gap ($E_g$), electronic sensitivity (%$\Delta E_g$), Fermi energy ($E_F$), adsorption energy ($E_{ads}$), BSSE-corrected adsorption energy, Grimme D3(BJ) dispersion contribution ($E_{disp}$) and dipole moment (μ) of free phosgene and the eight noble metal-doped adsorption complexes in the gas phase. %$\Delta E_g$ is relative to the corresponding bare doped cage of Table 1.

| System | *E*coh (eV/atom) | Bond length (Å) | $E_H$ (eV) | $E_L$ (eV) | $E_g$ (eV) | %$\Delta E_g$ | $E_F$ (eV) | $E_{ads}$ (kJ mol$^{-1}$) | $E_{ads}^{BSSE}$ (kJ mol$^{-1}$) | $E_{disp}$ (kJ mol$^{-1}$) | $\mu$ (D) |
|---|---|---|---|---|---|---|---|---|---|---|---|
| $COCl_2$ | −3.651 | -- | −9.10 | −2.08 | 7.02 | -- | −5.59 | -- | -- | -- | 1.22 |
| Ag–O | −5.331 | 2.34 (Ag···O) | −6.49 | −5.13 | 1.37 | 15.25 | −5.81 | −57.83 | −50.17 | −25.08 | 5.34 |
| Ag–Cl | −5.326 | 2.74 (Ag···Cl) | −6.67 | −5.35 | 1.33 | 11.83 | −6.01 | −43.38 | −35.45 | −30.74 | 3.49 |
| Au–O | −5.338 | 2.28 (Au···O) | −6.36 | −4.56 | 1.80 | 23.36 | −5.46 | −61.06 | −52.84 | −25.93 | 5.68 |
| Au–Cl | −5.323 | 2.96 (N···Cl) | −6.88 | −5.40 | 1.49 | 2.03 | −6.14 | −19.49 | −13.96 | −11.08 | 1.83 |
| Pd–O | −5.410 | 2.26 (Pd···O) | −6.47 | −3.96 | 2.51 | 21.70 | −5.22 | −67.19 | −58.70 | −26.75 | 5.74 |
| Pd–Cl | −5.406 | 2.58 (Pd···Cl) | −6.59 | −4.20 | 2.39 | 16.11 | −5.39 | −56.21 | −47.08 | −30.86 | 4.37 |
| Pt–O | −5.455 | 2.23 (Pt···O) | −6.08 | −3.59 | 2.49 | 18.42 | −4.84 | −69.09 | −60.71 | −27.64 | 5.88 |
| Pt–Cl | −5.452 | 2.64 (Pt···Cl) | −6.58 | −3.34 | 3.23 | 53.87 | −4.96 | −60.87 | −52.88 | −29.85 | 5.98 |

### 3.4 HOMO–LUMO and Density of States Analysis

Figure 3 shows the frontier molecular orbital (FMO) electron density maps and the total DOS of the four oxygen-side complexes, each DOS overlaid on that of the pristine B–O complex for reference; the chlorine-side set is given in Figure S3. Among the adsorbed configurations (Tables 1 and 2), B–Cl has the lowest HOMO energy at −8.06 eV, and Pt–O the highest HOMO at −6.08 eV. Au–Cl has the lowest LUMO of any complex (−5.40 eV), just below Ag–Cl (−5.35 eV), and is the deepest mid-gap state in the adsorbed series. Despite these differences in absolute orbital energies, adsorption widens the HOMO–LUMO gap of every doped cage relative to the corresponding bare cage, which is consistent with charge redistribution on binding. The density maps localise frontier orbital density on the dopant atoms and the neighbouring nitrogen atoms, indicating strong metal–adsorbate coupling, while the LUMO of every complex spreads onto the phosgene fragment.

The DOS spectra show the pristine $B_{12}N_{12}$ gap of 6.74 eV narrowing progressively on doping, reaching a minimum of 1.19 eV for $AgB_{11}N_{12}$. A new high-energy occupied state introduced by the dopant drives this narrowing [47]. After $COCl_2$ adsorption, Ag–O and Ag–Cl retain the smallest gaps (1.37 and 1.33 eV) and therefore the highest conductivity. Au-doped systems stay narrow as well, at 1.80 eV (Au–O) and 1.49 eV (Au–Cl). Pd- and Pt-doped systems mostly sit higher, at 2.39–2.51 eV, giving lower conductivity but a broader dynamic sensing range. Pt–Cl is a clear exception, widening to 3.23 eV (Figure S3d). This is far above every other doped configuration and consistent with its $\Delta E_g$ of +53.87% in Table 2. The DOS confirms the anomaly as a genuine electronic effect of the Pt–Cl geometry, not an artefact of the energy-gap calculation.

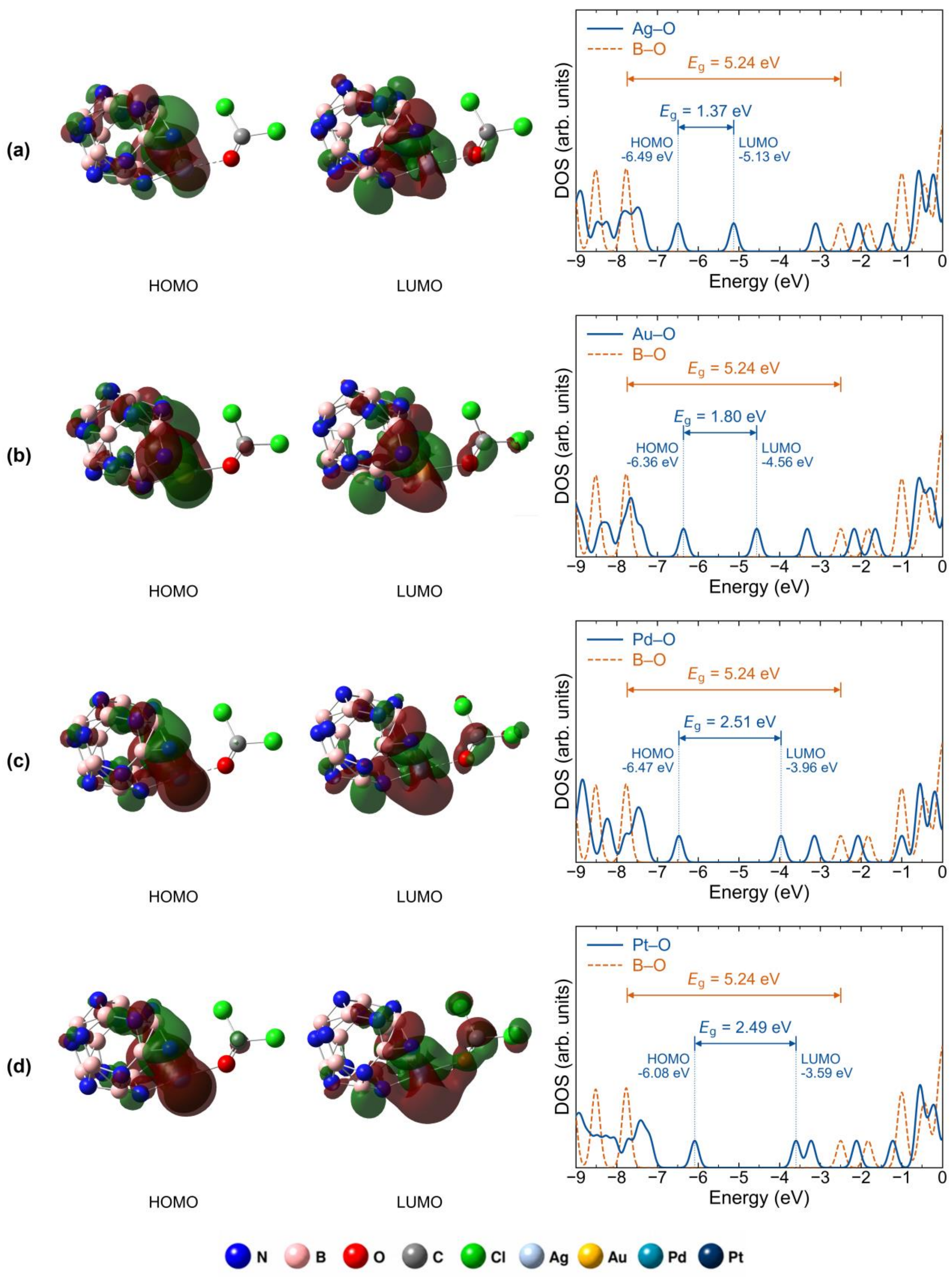

**Figure 3.** Representations of the HOMO, LUMO and DOS of the optimised geometry of $AgB_{11}N_{12}$ (a), $AuB_{11}N_{12}$ (b), $PdB_{11}N_{12}$ (c) and $PtB_{11}N_{12}$ (d) with phosgene gas ($COCl_2$) adsorbed near the oxygen side. The dashed orange trace in each DOS panel is the pristine B–O complex.

### 3.5 Adsorption Energy Analysis

Table 2 collects the adsorption energies without and with counterpoise correction, together with the Grimme D3(BJ) dispersion contribution, and Figure 4 compares the two sets graphically. The uncorrected values span −12.82 kJ mol$^{-1}$ (B–Cl, pristine) to −69.09 kJ mol$^{-1}$ (Pt–O). For each of the four metals the better of the two orientations binds $COCl_2$ far more strongly than either pristine complex, confirming that metal doping enhances binding. The one doped configuration that does not is Au–Cl, at −19.49 kJ mol$^{-1}$ marginally weaker than B–O (−24.90 kJ mol$^{-1}$), because its fragments never come into close contact. Across the doped set the values fall between −19.49 and −69.09 kJ mol$^{-1}$ (−0.202 to −0.716 eV), a narrow and consistent range. Counterpoise correction removes between 4 and 9 kJ mol$^{-1}$ and leaves the ordering unchanged apart from an exchange between Au–O and Pt–Cl, which differ by less than 0.2 kJ mol$^{-1}$; the corrected doped values run from −13.96 kJ mol$^{-1}$ (Au–Cl) to −60.71 kJ mol$^{-1}$ (Pt–O). These magnitudes indicate moderate, favourable adsorption rather than strong chemisorption, though the free-energy analysis in Section 3.14 shows that they overstate how many of these configurations actually bind at room temperature. The D3(BJ) dispersion term contributes between −11.08 kJ mol$^{-1}$ (Au–Cl, whose fragments never approach closely) and −30.86 kJ mol$^{-1}$ (Pd–Cl), that is roughly 40% to 70% of the total binding energy, so dispersion is indispensable at this level of theory. The uncorrected and counterpoise-corrected energies in Figure 4 track each other closely and preserve the same trend [48], [49].

This narrow window contrasts with the wide dispersion reported for the 3d series by Sousa and Varela Júnior [20], where several early metals bind far too strongly for practical sensor regeneration. The four noble metals instead occupy a moderate adsorption window suited to reversible sensing.

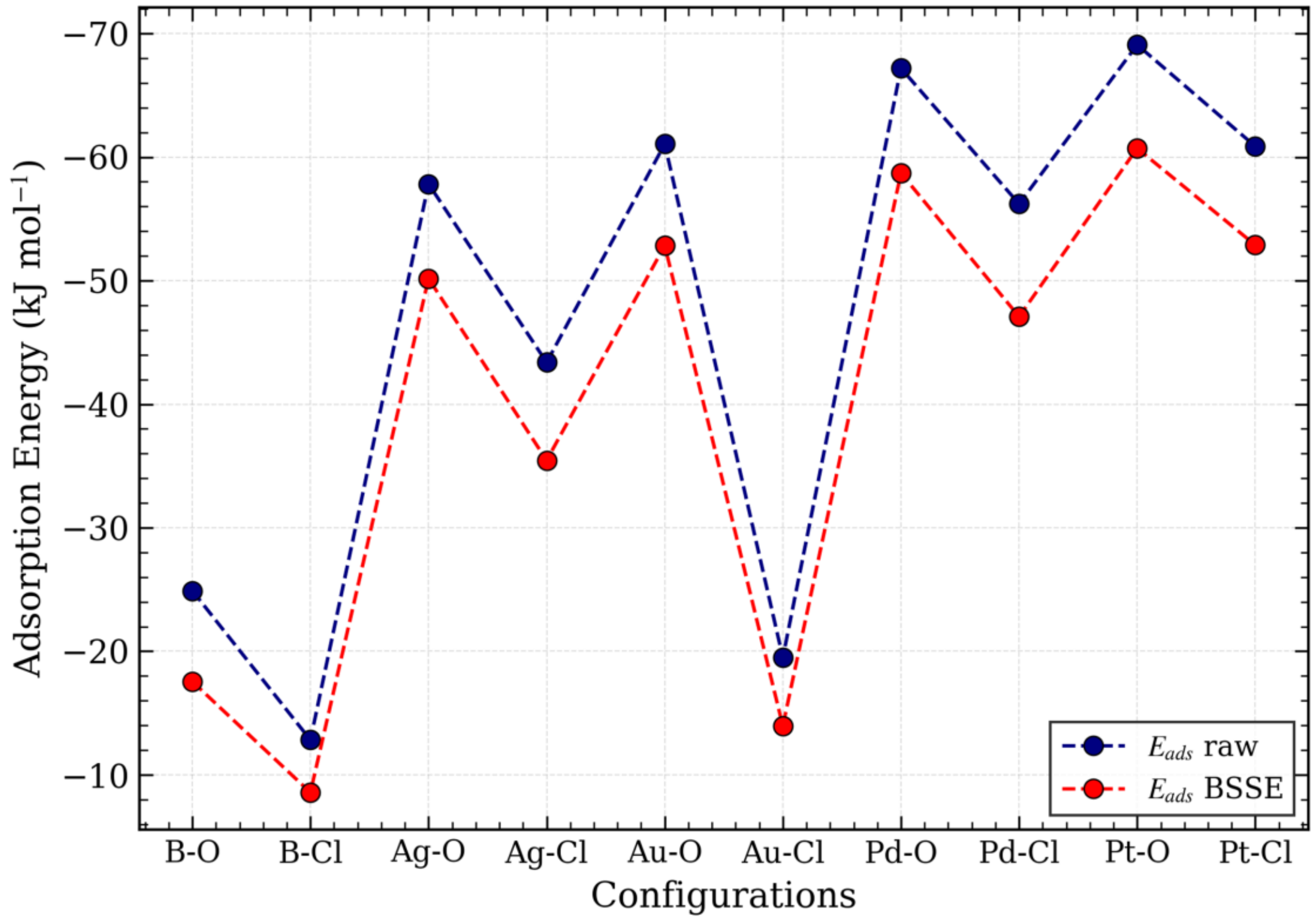

**Figure 4.** Adsorption energies of $COCl_2$ on the different configurations (raw versus BSSE-corrected) at the B3LYP-D3(BJ)/GENECP [LanL2DZ/6-311++G(d,p)] level.

### 3.6 Global Reactivity Indices

The global reactivity descriptors of Table 3 support the adsorption results above. IP falls from 7.97 eV in the pristine cage to 6.73 eV (Ag), 6.76 eV (Au), 6.64 eV (Pd), and 6.33 eV (Pt). EA rises from 1.23 eV to 5.55 eV for $AgB_{11}N_{12}$, reflecting the mid-gap states silver introduces. After adsorption, Au–Cl reaches the highest EA of any complex at 5.40 eV, narrowly ahead of Ag–Cl (5.35 eV); Au–Cl is also the weakest adsorber, so a high electron affinity alone does not guarantee strong binding. Chemical hardness drops sharply on doping, $AgB_{11}N_{12}$ reaching the minimum ($\eta$ = 0.59 eV) with softness S = 0.84 $eV^{-1}$, consistent with the high polarisability of silver. $\chi$ rises with doping to 6.14 eV for $AgB_{11}N_{12}$, and electrophilicity $\omega$ climbs from 3.14 eV in the pristine cage to 31.79 eV in $AgB_{11}N_{12}$, with Ag–Cl holding the highest post-adsorption value at 27.24 eV. For reference, free $COCl_2$ has IP = 9.10 eV, EA = 2.08 eV and $\omega$ = 4.45 eV, so doping lifts the electrophilicity of the cage far above that of the adsorbate. Ag-doped configurations are therefore the most electrophilic without being the strongest binders, matching their moderate adsorption energies in Table 2.

**Table 3.** Global reactivity descriptors — ionisation potential (IP), electron affinity (EA), chemical potential ($\mu$), electrophilicity index ($\omega$), chemical hardness ($\eta$), global softness (S) and electronegativity ($\chi$) — for all systems in the gas phase.

| System | IP (eV) | EA (eV) | $\mu$ (eV) | $\omega$ (eV) | $\eta$ (eV) | $S$ ($eV^{-1}$) | $\chi$ (eV) |
|---|---|---|---|---|---|---|---|
| $COCl_2$ | 9.10 | 2.08 | −5.59 | 4.45 | 3.51 | 0.14 | 5.59 |
| $B_{12}N_{12}$ | 7.97 | 1.23 | −4.60 | 3.14 | 3.37 | 0.15 | 4.60 |
| B–O | 7.74 | 2.50 | −5.12 | 5.00 | 2.62 | 0.19 | 5.12 |
| B–Cl | 8.06 | 1.98 | −5.02 | 4.14 | 3.04 | 0.16 | 5.02 |
| $AgB_{11}N_{12}$ | 6.73 | 5.55 | −6.14 | 31.79 | 0.59 | 0.84 | 6.14 |
| Ag–O | 6.49 | 5.13 | −5.81 | 24.70 | 0.68 | 0.73 | 5.81 |
| Ag–Cl | 6.67 | 5.35 | −6.01 | 27.24 | 0.66 | 0.75 | 6.01 |
| $AuB_{11}N_{12}$ | 6.76 | 5.30 | −6.03 | 24.93 | 0.73 | 0.69 | 6.03 |
| Au–O | 6.36 | 4.56 | −5.46 | 16.57 | 0.90 | 0.56 | 5.46 |
| Au–Cl | 6.88 | 5.40 | −6.14 | 25.33 | 0.74 | 0.67 | 6.14 |
| $PdB_{11}N_{12}$ | 6.64 | 4.58 | −5.61 | 15.27 | 1.03 | 0.49 | 5.61 |
| Pd–O | 6.47 | 3.96 | −5.22 | 10.85 | 1.25 | 0.40 | 5.22 |
| Pd–Cl | 6.59 | 4.20 | −5.39 | 12.15 | 1.20 | 0.42 | 5.39 |
| $PtB_{11}N_{12}$ | 6.33 | 4.23 | −5.28 | 13.27 | 1.05 | 0.48 | 5.28 |
| Pt–O | 6.08 | 3.59 | −4.84 | 9.41 | 1.24 | 0.40 | 4.84 |
| Pt–Cl | 6.58 | 3.34 | −4.96 | 7.61 | 1.62 | 0.31 | 4.96 |

### 3.7 Charge Transfer Analysis

Table 4 lists Mulliken and natural (NBO) fragment charges for every system. The charge transfer is defined as $\Delta q = q(COCl_2)$, the net charge carried by the adsorbate in the complex, so a positive $\Delta q$ means that phosgene is electron-poor in the complex and has donated electron density to the cage.

Both partitions agree that the transfer is small and directed from phosgene to the cage in every O-down configuration. The NBO values are 0.084 e (B–O), 0.099 e (Ag–O), 0.132 e (Au–O), 0.119 e (Pd–O) and 0.132 e (Pt–O). The Cl-down configurations span a wider range: 0.124 e (Ag–Cl), 0.176 e (Pd–Cl) and 0.331 e (Pt–Cl), while B–Cl (−0.006 e) and Au–Cl (−0.002 e) are essentially neutral, consistent with their long fragment separations and weak binding. Pt–Cl carries by far the largest transfer of the series, matching its two Pt···Cl bond paths and its anomalously large gap change. The Mulliken values are systematically larger and less consistent between related systems, as expected of a basis-set-dependent partition, but they reproduce the same direction and the same extremes. Even the largest transfer, 0.331 e, is an order of magnitude below the one-electron scale of a chemical bond, which places all of these interactions firmly in the physisorption regime and agrees with the QTAIM classification in Section 3.13.

The NBO charge on the dopant itself changes little on adsorption. It falls from +0.929 e to +0.904 e in Ag–O and from +0.912 e to +0.880 e in Pt–O, so the electron density that phosgene donates is delocalised across the cage rather than accumulating on the metal.

**Table 4.** Mulliken and NBO fragment charges and the resulting charge transfer on $COCl_2$ adsorption.

| Configuration | $q$(X) Mulliken (e) | $q$(cage) Mulliken (e) | $\Delta q = q(COCl_2)$ Mulliken (e) | $q$(X) NBO (e) | $q$(cage) NBO (e) | $\Delta q = q(COCl_2)$ NBO (e) |
|---|---|---|---|---|---|---|
| ***Free adsorbate*** | | | | | | |
| $COCl_2$ | -- | 0.000 | 0.000 | -- | 0.000 | 0.000 |
| ***Undoped*** | | | | | | |
| $B_{12}N_{12}$ | -- | 0.000 | -- | -- | 0.000 | -- |
| B–O (O-down) | -- | −0.063 | 0.063 | -- | −0.084 | 0.084 |
| B–Cl (Cl-down) | -- | −0.075 | 0.075 | -- | 0.005 | −0.006 |
| ***Ag-doped*** | | | | | | |
| $AgB_{11}N_{12}$ | 0.218 | −0.218 | -- | 0.929 | −0.929 | -- |
| Ag–O (O-down) | 0.092 | −0.273 | 0.181 | 0.904 | −1.003 | 0.099 |
| Ag–Cl (Cl-down) | −0.037 | −0.283 | 0.320 | 0.843 | −0.967 | 0.124 |
| ***Au-doped*** | | | | | | |
| $AuB_{11}N_{12}$ | 0.446 | −0.446 | -- | 0.949 | −0.949 | -- |
| Au–O (O-down) | 0.250 | −0.532 | 0.282 | 0.952 | −1.084 | 0.132 |
| Au–Cl (Cl-down) | 0.536 | −0.428 | −0.108 | 0.958 | −0.956 | −0.002 |
| ***Pd-doped*** | | | | | | |
| $PdB_{11}N_{12}$ | 0.154 | −0.154 | -- | 0.852 | −0.852 | -- |
| Pd–O (O-down) | 0.033 | −0.224 | 0.191 | 0.798 | −0.917 | 0.119 |
| Pd–Cl (Cl-down) | −0.237 | −0.049 | 0.286 | 0.704 | −0.880 | 0.176 |
| ***Pt-doped*** | | | | | | |
| $PtB_{11}N_{12}$ | 0.687 | −0.687 | -- | 0.912 | −0.912 | -- |
| Pt–O (O-down) | 0.559 | −0.800 | 0.241 | 0.880 | −1.012 | 0.132 |
| Pt–Cl (Cl-down) | 0.097 | −0.784 | 0.687 | 0.537 | −0.868 | 0.331 |

$\Delta q = q(COCl_2)$ is the net charge on the adsorbate; a positive value means that phosgene donates electron density to the cage. For the NBO method and its basis-set stability see A. E. Reed, R. B. Weinstock and F. Weinhold, J. Chem. Phys. 83 (1985) 735, and A. E. Reed, L. A. Curtiss and F. Weinhold, Chem. Rev. 88 (1988) 899.

## 3.8 Bond Length Analysis

Table 5 collects bond lengths for all configurations. For the pristine cage the C=O bond is 1.18 Å (B–O) and 1.17 Å (B–Cl), against 1.17 Å in free phosgene, with adsorption energies of −24.90 and −12.82 kJ mol$^{-1}$. Noble metal doping elongates C=O and shortens the metal–O contact: Ag–O, Au–O, Pd–O, and Pt–O all extend C=O to 1.19–1.20 Å, with metal···O distances of 2.23–2.34 Å and $E_{ads}$ from −57.83 to −69.09 kJ mol$^{-1}$. Pt–O has the tightest contact (Pt···O = 2.23 Å) and the strongest binding. The elongation reflects charge transfer from the metal into the carbonyl π* antibonding orbital, which weakens C=O; the QTAIM analysis of Table S2 confirms this independently. Chlorine-facing geometries give weaker and more variable metal···Cl contacts. Ag···Cl is 2.74 Å and Pd···Cl 2.58 Å, while Pt–Cl is bidentate, with two nearly equal Pt···Cl contacts of 2.64 and 2.66 Å. Au–Cl is the extreme case: its Au···Cl separation of 3.92 Å is essentially non-bonded, and the closest approach between the two fragments is instead an N···Cl contact of 2.96 Å on the cage surface away from the dopant. QTAIM finds no Au···Cl bond critical point at all, only that N···Cl path, which explains why Au–Cl is the weakest doped adsorber overall. The B–N ring bonds provide an independent structural check: they hold at 1.47 Å in every doped cage and every complex, close to the pristine $b_{66}$ and $b_{64}$ values of 1.436 and 1.484 Å, so metal substitution and adsorption leave the cage framework essentially intact. The dopant itself sits 2.55–2.81 Å from its boron neighbours and 2.05–2.23 Å from its nitrogen neighbours in the bare cages, and adsorption moves these distances by no more than 0.16 Å, so the metal is held firmly in the framework. The C–Cl bond, finally, mirrors the C=O response in reverse. It contracts to 1.72–1.73 Å in the oxygen-facing complexes and stretches to 1.79–1.84 Å in the three chlorine-facing complexes that make genuine contact, against 1.76 Å in free phosgene; only Au–Cl stays at the free-molecule value. Whichever atom faces the dopant is therefore the one whose bond to carbon lengthens, which is exactly what donation into the corresponding antibonding orbital predicts.

**Table 5.** Selected bond lengths (Å) for free phosgene, the bare cages and all adsorption configurations.

| Cage | Adsorption mode | X | X–B (Å) | X–N (Å) | B–N (Å) | X···Cl₂ (Å) | X···Cl1 (Å) | X···O (Å) | C–Cl₂ (Å) | C=O (Å) | Shortest contact (Å) |
|---|---|---|---|---|---|---|---|---|---|---|---|
| $COCl_2$ | free molecule | -- | -- | -- | -- | -- | -- | -- | 1.76 | 1.17 | -- |
| $B_{12}N_{12}$ | bare cage | B | -- | -- | 1.47 | -- | -- | -- | -- | -- | -- |
| $B_{12}N_{12}$ | O-down | B | -- | 1.48 | 1.47 | 3.50 | 4.84 | 2.34 | 1.75 | 1.18 | 2.34 (B···O) |
| $B_{12}N_{12}$ | Cl-down | B | -- | 1.47 | 1.47 | 3.81 | 5.96 | 6.42 | 1.76 | 1.17 | 3.05 (N···Cl) |
| $AgB_{11}N_{12}$ | bare cage | Ag | 2.81 | 2.23 | 1.47 | -- | -- | -- | -- | -- | -- |
| $AgB_{11}N_{12}$ | O-down | Ag | 2.75 | 2.22 | 1.47 | 3.63 | 4.83 | 2.34 | 1.73 | 1.19 | 2.34 (Ag···O) |
| $AgB_{11}N_{12}$ | Cl-down | Ag | 2.75 | 2.21 | 1.47 | 2.74 | 4.42 | 3.73 | 1.82 | 1.17 | 2.74 (Ag···Cl) |
| $AuB_{11}N_{12}$ | bare cage | Au | 2.67 | 2.14 | 1.47 | -- | -- | -- | -- | -- | -- |
| $AuB_{11}N_{12}$ | O-down | Au | 2.67 | 2.18 | 1.47 | 3.69 | 4.80 | 2.28 | 1.72 | 1.20 | 2.28 (Au···O) |
| $AuB_{11}N_{12}$ | Cl-down | Au | 2.67 | 2.14 | 1.47 | 3.92 | 6.85 | 5.70 | 1.76 | 1.18 | 2.96 (N···Cl) |
| $PdB_{11}N_{12}$ | bare cage | Pd | 2.55 | 2.06 | 1.47 | -- | -- | -- | -- | -- | -- |
| $PdB_{11}N_{12}$ | O-down | Pd | 2.61 | 2.10 | 1.47 | 3.71 | 4.80 | 2.26 | 1.73 | 1.19 | 2.26 (Pd···O) |
| $PdB_{11}N_{12}$ | Cl-down | Pd | 2.57 | 2.08 | 1.47 | 2.58 | 3.80 | 4.19 | 1.84 | 1.17 | 2.58 (Pd···Cl) |
| $PtB_{11}N_{12}$ | bare cage | Pt | 2.57 | 2.05 | 1.47 | -- | -- | -- | -- | -- | -- |
| $PtB_{11}N_{12}$ | O-down | Pt | 2.61 | 2.09 | 1.47 | 3.74 | 4.77 | 2.23 | 1.72 | 1.20 | 2.23 (Pt···O) |
| $PtB_{11}N_{12}$ | Cl-down | Pt | 2.73 | 2.18 | 1.47 | 2.64 | 2.66 | 4.39 | 1.79 | 1.16 | 2.64 (Pt···Cl) |

The final column gives the shortest cage···adsorbate contact and the atom pair that forms it.

### 3.9 Dipole Moments

Tables 1 and 2 report gas-phase dipole moments for pristine $B_{12}N_{12}$, the bare doped cages and the $COCl_2$ complexes, and Table 7 adds the aqueous values. The pristine cage has a dipole moment of 0.00 D by symmetry. Adsorption raises it to 2.56 D (B–O) and 1.65 D (B–Cl), a difference of about 0.9 D that reflects the change in adsorption orientation. Noble metal doping breaks the cage symmetry and introduces dipole moments of 1.52–2.94 D. Adsorption raises these further, to 1.83–5.98 D in the gas phase.

Ordered by gas-phase dipole moment following $COCl_2$ adsorption (from Table 2): Pt–Cl (5.98 D) > Pt–O (5.88 D) > Pd–O (5.74 D) > Au–O (5.68 D) > Ag–O (5.34 D) > Pd–Cl (4.37 D) > Ag–Cl (3.49 D) > Au–Cl (1.83 D). For silver, gold and palladium the oxygen-facing configuration carries the larger dipole, reflecting the stronger charge redistribution at the metal–O contact; only for platinum are the two orientations comparable, with Pt–Cl marginally ahead, which is consistent with its bidentate Pt···Cl geometry and the large charge transfer reported in Table 4. Au–Cl is the clear outlier at 1.83 D, barely above the 1.22 D of free phosgene, as expected for a complex whose fragments never come into close contact. Solvent effects on these values are covered in Section 3.12.

### 3.10 Molecular Electrostatic Potential Analysis

MEP maps of the pristine cage, free phosgene and the two pristine complexes are shown in Figure 1(b, d, f, h), those of the oxygen-side doped complexes in Figure 2 ($a_3$–$d_3$) and those of the chlorine-side complexes in Figure S2 ($a_3$–$d_3$). In the pristine cage, electron-deficient (blue) regions surround boron atoms and electron-rich (yellow/red) regions surround nitrogen atoms. Doping shifts the electron-deficient region onto the metal centre, creating a favourable electrostatic site for $COCl_2$. The carbonyl oxygen carries a partial negative charge, so its approach to that site is electrostatically favourable. Figure 5 makes the mechanism explicit for the silver system: the ESP of the isolated $AgB_{11}N_{12}$ cage, of the isolated $COCl_2$ molecule and of the optimised Ag–O complex show the electron-rich carbonyl oxygen docking onto the electron-poor silver centre, with the MEP of the complex clearly redistributed between the metal and $COCl_2$. The same density redistribution appears in the MEP maps of every doped complex and is consistent with the charge transfer of Table 4 and with the QTAIM analysis of Section 3.13. This picture matches the bond lengths in Table 5: the O-down complexes sit closest to the electron-deficient metal centre (X···O = 2.23–2.34 Å) and bind most strongly, whereas Au–Cl never brings its chlorine into that region (Au···Cl = 3.92 Å), makes contact only through a distant N···Cl approach on the far side of the cage, and is the weakest binder in the series.

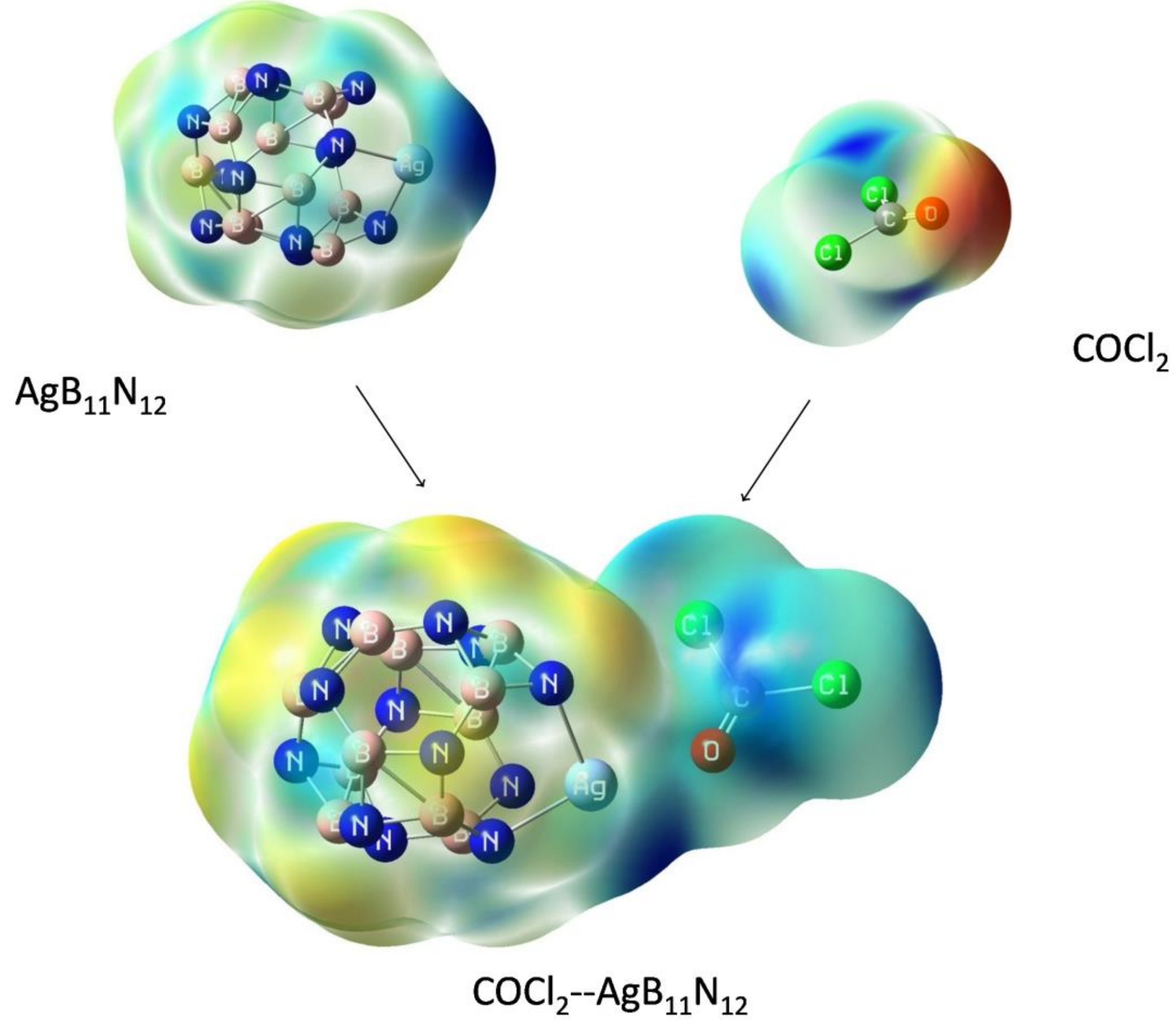

**Figure 5.** MEP surface analysis illustrating the interaction and adsorption behaviour of phosgene ($COCl_2$) on the $AgB_{11}N_{12}$ nanocluster: ESP distribution of the pristine $AgB_{11}N_{12}$ cluster, the isolated $COCl_2$ molecule, and the optimised $COCl_2$–$AgB_{11}N_{12}$ adsorption complex.

### 3.11 Work Function, Fermi Level and Recovery Time

Sensor performance is assessed in this work through four quantities: the energy gap change $\Delta E_g$, which carries the electrical signal; the work function Φ, which governs field-effect and thermionic transduction; the free energy of adsorption $\Delta G_{qh}$, which decides whether the analyte binds at all; and the desorption temperature $T_{des}$ at which that free energy changes sign. Table 6 collects the electronic quantities together with the Arrhenius recovery time; the thermodynamic quantities are treated in Sections 3.14 and 3.15. The work function was computed as $\Phi = -E_F$ (Eq. 13). The pristine nanocage gives Φ = 4.60 eV, and doping raises it, reaching 6.14 eV for $AgB_{11}N_{12}$.

Work function changes on $COCl_2$ adsorption vary widely. Ag–O falls by 5.37% to 5.81 eV, Au–O shows the largest decrease (−9.45%), and Au–Cl rises slightly (+1.83%). Such shifts in Φ are exploitable through the Richardson–Dushman relation, which links the thermionic emission current density j to the work function [50]:

$$j = AT^2 \exp\left(-\frac{\Phi}{\kappa T}\right) \tag{17}$$

where A is the Richardson constant ($1.20 \times 10^6$ A m−2 $K^{-2}$), κ is the Boltzmann constant, and T is the temperature.

The recovery time τ, the mean residence time of an adsorbed molecule before it desorbs thermally, was estimated from transition-state theory (Arrhenius desorption):

$$\tau = \nu_0^{-1} \exp\left(\frac{|E_{\mathrm{ads}}|}{k_B T}\right) \tag{18}$$

with attempt frequency $\nu_0 = 10^{12}$ $s^{-1}$ and T = 298.15 K. A residence time is meaningful only for a configuration that is retained at that temperature, so τ is reported for the six complexes with $\Delta G_{qh} \leq +2$ kJ $mol^{-1}$ (Section 3.14) and omitted for B–O, B–Cl, Ag–Cl and Au–Cl, which do not bind phosgene at 298 K. With the counterpoise-corrected adsorption energies, τ rises from 0.18 ms (Pd–Cl) and 0.62 ms (Ag–O) through 1.8 ms (Au–O, Pt–Cl) to 19 ms (Pd–O) and 43 ms (Pt–O) (Table 6); the spread follows from the exponential, each additional 0.059 eV of binding lengthening τ tenfold. Even Pt–O releases phosgene within a few hundredths of a second, so recovery is never rate-limiting. Recovery times of $10^{-2}$–$10^2$ s are typical of transition-metal-modified $B_{12}N_{12}$ sensors evaluated with the same expression and constants, the longer values belonging to chemisorbed CO or $H_2$ on 3d- or 4d-metal sites with $|E_{ads}| \approx 0.7$–1.0 eV [15], whereas the early 3d metals that bind phosgene by more than 5 eV [20] would not release it on any practical timescale; the noble-metal cages therefore lie at the fast, fully reversible end of that range. Because τ neglects the entropy of desorption it describes how quickly a bound molecule leaves, not whether the cage retains phosgene; that is decided by $\Delta G_{qh}$ and $T_{des}$ (Section 3.15). DFT also overestimates $|E_{ads}|$ by 20–40%, so the real recovery times will be shorter still.

**Table 6.** Fermi level ($E_F$), electronic sensitivity (%$\Delta E_g$), work function ($\Phi$) and its change (%$\Delta\Phi$), with the counterpoise-corrected adsorption energy $|E_{ads}|$ and the corresponding Arrhenius recovery time τ at 298.15 K, in the gas phase.

| System | *EF* (eV) | %Δ*Eg* | Φ (eV) | %ΔΦ | \|*Eads*\| (eV) | *τ (ms)* |
|---|---|---|---|---|---|---|
| ***Pristine and doped nanocages*** | | | | | | |
| $B_{12}N_{12}$ | −4.60 | -- | 4.60 | -- | -- | -- |
| $AgB_{11}N_{12}$ | −6.14 | -- | 6.14 | -- | -- | -- |
| $AuB_{11}N_{12}$ | −6.03 | -- | 6.03 | -- | -- | -- |
| $PdB_{11}N_{12}$ | −5.61 | -- | 5.61 | -- | -- | -- |
| $PtB_{11}N_{12}$ | −5.28 | -- | 5.28 | -- | -- | -- |
| ***Undoped complexes*** | | | | | | |
| B–O | −5.12 | −22.24 | 5.12 | 11.32 | 0.182 | -- |
| B–Cl | −5.02 | −9.85 | 5.02 | 9.05 | 0.089 | -- |
| ***Ag-doped*** | | | | | | |
| Ag–O | −5.81 | 15.25 | 5.81 | −5.37 | 0.520 | 0.616 |
| Ag–Cl | −6.01 | 11.83 | 6.01 | −2.11 | 0.367 | -- |
| ***Au-doped*** | | | | | | |
| Au–O | −5.46 | 23.36 | 5.46 | −9.45 | 0.548 | 1.83 |
| Au–Cl | −6.14 | 2.03 | 6.14 | 1.83 | 0.145 | -- |
| ***Pd-doped*** | | | | | | |
| Pd–O | −5.22 | 21.70 | 5.22 | −6.99 | 0.608 | 18.9 |
| Pd–Cl | −5.39 | 16.11 | 5.39 | −3.86 | 0.488 | 0.177 |
| ***Pt-doped*** | | | | | | |
| Pt–O | −4.84 | 18.42 | 4.84 | −8.35 | 0.629 | 42.9 |
| Pt–Cl | −4.96 | 53.87 | 4.96 | −6.07 | 0.548 | 1.83 |

$\tau = \nu_0^{-1} \exp(|E_{ads}|/k_BT)$ is the recovery time from transition-state theory (Eq. 18), with attempt frequency $\nu_0 = 10^{12}$ s$^{-1}$ and T = 298.15 K, evaluated with the counterpoise-corrected adsorption energies; it is given only for the configurations that are retained at 298 K ($\Delta G_{qh} \leq +2$ kJ mol$^{-1}$, Table 8) and omitted for those that do not bind.

### 3.12 Solvent Effects

Solvent stability was evaluated with the CPCM continuum model in water, as single-point calculations on the gas-phase geometries; Table 7 and Table S1 collect the results. All solvation energies ($E_{sol}$) are negative, so every cage and every complex is stabilised in water. Among the bare cages $PtB_{11}N_{12}$ is the most strongly solvated by a wide margin (−122.51 kJ mol$^{-1}$), ahead of $PdB_{11}N_{12}$ (−91.64), $AgB_{11}N_{12}$ (−71.28) and $AuB_{11}N_{12}$ (−50.45 kJ mol$^{-1}$); pristine $B_{12}N_{12}$, which has no dipole, gains only −17.58 kJ mol$^{-1}$, and free $COCl_2$ only −6.45 kJ mol$^{-1}$. That ranking follows the aqueous dipole moments of Table 7 exactly (6.60 D for Pt, 6.22 for Pd, 6.01 for Ag and 4.07 D for Au), as expected of a reaction-field model, and it agrees with the cohesive-energy ordering of Table 1 except that silver and gold, which differ by only 0.006 eV per atom, exchange places. Platinum is therefore the most stable choice by cohesion, solvation and adsorption strength alike. Dipole moments rise sharply in water: every doped cage at least doubles its gas-phase value and $PtB_{11}N_{12}$ more than quadruples it, from 1.52 to 6.60 D, while among the complexes Pt–Cl reaches 9.64 D and Pt–O 9.53 D, the two highest of any system. Table 7 also shows how the frontier orbitals themselves respond: solvation raises the HOMO of every system except B–O, most strongly for the platinum systems, the HOMO of $PtB_{11}N_{12}$ moving from −6.33 eV in the gas phase to −5.35 eV in water. The LUMO rises by slightly more in most cases, so the aqueous gaps come out marginally wider than the gas-phase ones for every system except B–Cl and Pd–Cl.

Because the bare doped cages are solvated more strongly than their complexes, aqueous adsorption energies are uniformly weaker than gas-phase ones. Water reduces $E_{ads}$ from −57.8 to −34.2 kJ mol$^{-1}$ for Ag–O, from −67.2 to −38.7 for Pd–O and from −69.1 to −31.1 for Pt–O, while Au–O is the least affected of the O-down set, falling only from −61.1 to −51.5 kJ mol$^{-1}$ and so becoming the strongest aqueous adsorber of the series. The corresponding aqueous recovery times at 298 K (Table S1) are all far below a second, ranging from 1.10 × $10^{-7}$ ms for B–Cl to 1.07 ms for Au–O, shorter than the corresponding gas-phase values of Table 6, so desorption in a humid environment is faster still.

Pt–Cl is the single exception. Its aqueous adsorption energy is +3.7 kJ $mol^{-1}$, that is marginally unbound, so no recovery time can be quoted. The cause is traceable in Table S1: $PtB_{11}N_{12}$ gains −122.51 kJ $mol^{-1}$ on solvation while the Pt–Cl complex gains only −64.42 kJ $mol^{-1}$, a difference of 58 kJ $mol^{-1}$. The corresponding difference for Pt–O is 31 kJ $mol^{-1}$, and for every other dopant the O-down and Cl-down complexes are solvated within 5 kJ $mol^{-1}$ of each other. The Pt–Cl solvation energy is therefore the outlier, not its gas-phase binding, which at −60.87 kJ $mol^{-1}$ is entirely normal. Wavefunction stability analysis on all three platinum CPCM wavefunctions returned stable solutions with unchanged energies, so the result is not an SCF artefact. It is best read as a limitation of applying a continuum model to a single fixed gas-phase geometry rather than as evidence that Pt–Cl cannot form in water.

Solvation also damps the sensing signal. $|\Delta E_g|$ shrinks in water for every doped system: Au–O falls from 23.36% to 17.35%, Pd–O from 21.70% to 18.53%, Pd–Cl from 16.11% to 9.87%, Ag–O from 15.25% to 12.00%, and Au–Cl effectively vanishes, from +2.03% to −0.60%. The Pt–Cl anomaly survives solvation, however: its gas-phase $\Delta E_g$ of +53.87% remains +44.27% in water, still more than twice that of any other configuration, so the large gap change is an intrinsic feature of the Pt–Cl geometry and not a gas-phase artefact. Overall, the doped systems remain more stable than the pristine cage in both phases, and a humid or aqueous operating environment damps but does not remove the sensing signal identified in the gas phase.

**Table 7.** Frontier-orbital properties and dipole moments of the nanocages and their $COCl_2$ complexes in the gas phase and in water.

| System | *E*H aq (eV) | *E*L aq (eV) | *E*g gas (eV) | *E*g aq (eV) | %Δ*E*g gas | %Δ*E*g aq | *μ* gas (D) | *μ* aq (D) |
|---|---|---|---|---|---|---|---|---|
| ***Free adsorbate*** | | | | | | | | |
| $COCl_2$ | −9.002 | −1.980 | 7.020 | 7.022 | -- | -- | 1.22 | 1.53 |
| ***Undoped*** | | | | | | | | |
| $B_{12}N_{12}$ | −7.943 | −1.127 | 6.744 | 6.816 | -- | -- | 0.00 | 0.00 |
| B–O | −7.826 | −2.243 | 5.244 | 5.583 | −22.24 | −18.09 | 2.56 | 3.07 |
| B–Cl | −7.949 | −1.931 | 6.080 | 6.018 | −9.85 | −11.72 | 1.65 | 1.87 |
| ***Ag-doped*** | | | | | | | | |
| $AgB_{11}N_{12}$ | −6.397 | −5.122 | 1.185 | 1.274 | -- | -- | 2.94 | 6.01 |
| Ag–O | −6.397 | −4.970 | 1.366 | 1.427 | 15.25 | 12.00 | 5.34 | 7.59 |
| Ag–Cl | −6.481 | −5.104 | 1.325 | 1.377 | 11.83 | 8.09 | 3.49 | 5.69 |
| ***Au-doped*** | | | | | | | | |
| $AuB_{11}N_{12}$ | −6.418 | −4.821 | 1.458 | 1.596 | -- | -- | 1.87 | 4.07 |
| Au–O | −6.268 | −4.395 | 1.799 | 1.874 | 23.36 | 17.35 | 5.68 | 7.90 |
| Au–Cl | −6.411 | −4.824 | 1.488 | 1.587 | 2.03 | −0.60 | 1.83 | 2.55 |
| ***Pd-doped*** | | | | | | | | |
| $PdB_{11}N_{12}$ | −6.018 | −3.847 | 2.061 | 2.172 | -- | -- | 2.24 | 6.22 |
| Pd–O | −6.157 | −3.583 | 2.508 | 2.574 | 21.70 | 18.53 | 5.74 | 8.77 |
| Pd–Cl | −6.156 | −3.770 | 2.393 | 2.386 | 16.11 | 9.87 | 4.37 | 7.94 |
| ***Pt-doped*** | | | | | | | | |
| $PtB_{11}N_{12}$ | −5.350 | −3.101 | 2.101 | 2.249 | -- | -- | 1.52 | 6.60 |
| Pt–O | −5.535 | −2.934 | 2.488 | 2.601 | 18.42 | 15.67 | 5.88 | 9.53 |
| Pt–Cl | −6.194 | −2.950 | 3.232 | 3.245 | 53.87 | 44.27 | 5.98 | 9.64 |

### 3.13 Bader AIM Theory and Non-Covalent Interaction Analysis

Bader's atoms-in-molecules (AIM) theory [51], [52] characterises the bonding at the metal–adsorbate interface. Bond critical points (BCPs) mark the saddle points of the electron density along the bond paths that link pairs of nuclei. Each BCP is described by its electron density ($\rho_b$), its Laplacian ($\nabla^2\rho_b$), and the kinetic, potential and total energy densities $G_b$, $V_b$ and $H_b = G_b + V_b$. These are linked by the local form of the virial theorem, which in atomic units reads:

$$\frac{1}{4}\nabla^2\rho_b = 2G_b + V_b \tag{19}$$

A small $\rho_b$ together with a positive $\nabla^2\rho_b$ marks a closed-shell interaction, either electrostatic or dispersive. The sign of $H_b$ refines that picture: $H_b \geq 0$ is purely closed-shell (CS); $H_b < 0$ with $\nabla^2\rho_b > 0$ indicates an intermediate interaction (I) showing the first signs of covalency; and $\nabla^2\rho_b < 0$ with $H_b < 0$ is a shared, covalent interaction (S). The ratio $|V_b|/G_b$ carries the same information, lying below 1 for closed-shell, between 1 and 2 for intermediate, and above 2 for shared interactions. Table S2 lists these parameters for every bond critical point in the eight doped complexes, computed directly from the B3LYP-D3(BJ) wavefunctions. Figure 6 shows the molecular graphs of the four oxygen-side complexes and Figure S5 those of the chlorine-side complexes; in each graph the cage···adsorbate bond critical point is circled and annotated with its distance, $\rho_b$, $\nabla^2\rho_b$, $H_b$ and interaction type, and the panel header gives the corresponding Espinosa interaction energy, so that the classification discussed below can be read directly from the figure.

Every cage···adsorbate bond critical point has a small $\rho_b$ (0.008–0.058 a.u.), a positive $\nabla^2\rho_b$, and $|V_b|/G_b$ well below 2. None is shared, so the bonding is closed-shell or intermediate throughout, consistent with the moderate adsorption energies of Table 2 rather than with covalent chemisorption. The metal···O density ranks Pt–O (0.0577) > Au–O (0.0553) > Pd–O (0.0488) > Ag–O (0.0428), and the metal···Cl density ranks Pt–Cl (0.0445 and 0.0433 for its two bond paths) > Pd–Cl (0.0427) > Ag–Cl (0.0316). The chlorine-facing series reproduces the counterpoise-corrected adsorption energies of Table 2 exactly; the oxygen-facing series agrees on the two extremes, platinum strongest and silver weakest, and exchanges gold and palladium, which differ by less than 6 kJ mol$^{-1}$. Au–Cl is the clearest case of all: no Au···Cl bond critical point exists, and the only path linking the two fragments is a very weak N13···Cl26 contact at 2.961 Å ($\rho_b = 0.0129$, $H_b = +0.0017$ a.u.), which is why Au–Cl is the weakest adsorbing configuration in the doped series (Figure S5b). Pt–Cl is the opposite extreme and the only bidentate contact in the set, with two Pt···Cl bond paths of 2.641 and 2.657 Å (Figure S5d). The Espinosa estimate $E_{int} = V_b/2$ gives −114.7 kJ mol$^{-1}$ for Pt···O, −100.7 for Au···O, −95.7 for Pd···O and −74.6 for Ag···O, and −65.7 and −63.2 kJ mol$^{-1}$ for the two Pt···Cl paths. These are larger in magnitude than the supermolecular adsorption energies, as expected of a local pairwise estimate that neglects the deformation cost of the fragments. They rank the four oxygen-facing contacts in the same order as $\rho_b$, Pt > Au > Pd > Ag, rather than in the order of the adsorption energies. Finally, the C=O bond critical point itself responds to the binding mode: its density falls from 0.442–0.453 a.u. in the chlorine-facing complexes to 0.417–0.424 a.u. in the oxygen-facing complexes, and its Laplacian falls from +0.16 to +0.29 a.u. down to +0.02 to +0.04 a.u. Both changes confirm charge transfer into the C=O $\pi^*$ orbital and the resulting bond weakening, the same mechanism behind the O-down preference seen throughout this work.

One caveat applies to the boundary between the intermediate and closed-shell classes. The two silver contacts fall on opposite sides of it by a very small margin: Ag···O25 has $H_b$ = 0.00000 a.u. and Ag···Cl26 has $H_b$ = −0.00092 a.u., so the first is classified closed-shell and the second intermediate even though the two interactions are chemically alike. That distinction should not be over-interpreted. Both are best read as closed-shell contacts at the threshold of incipient covalency, and the same caution applies to Pd···O26 ($H_b$ = −0.0019 a.u.) and Pt···O26 (−0.0024 a.u.), and indeed to any contact whose $H_b$ lies within a few times $10^{-3}$ a.u. of zero.

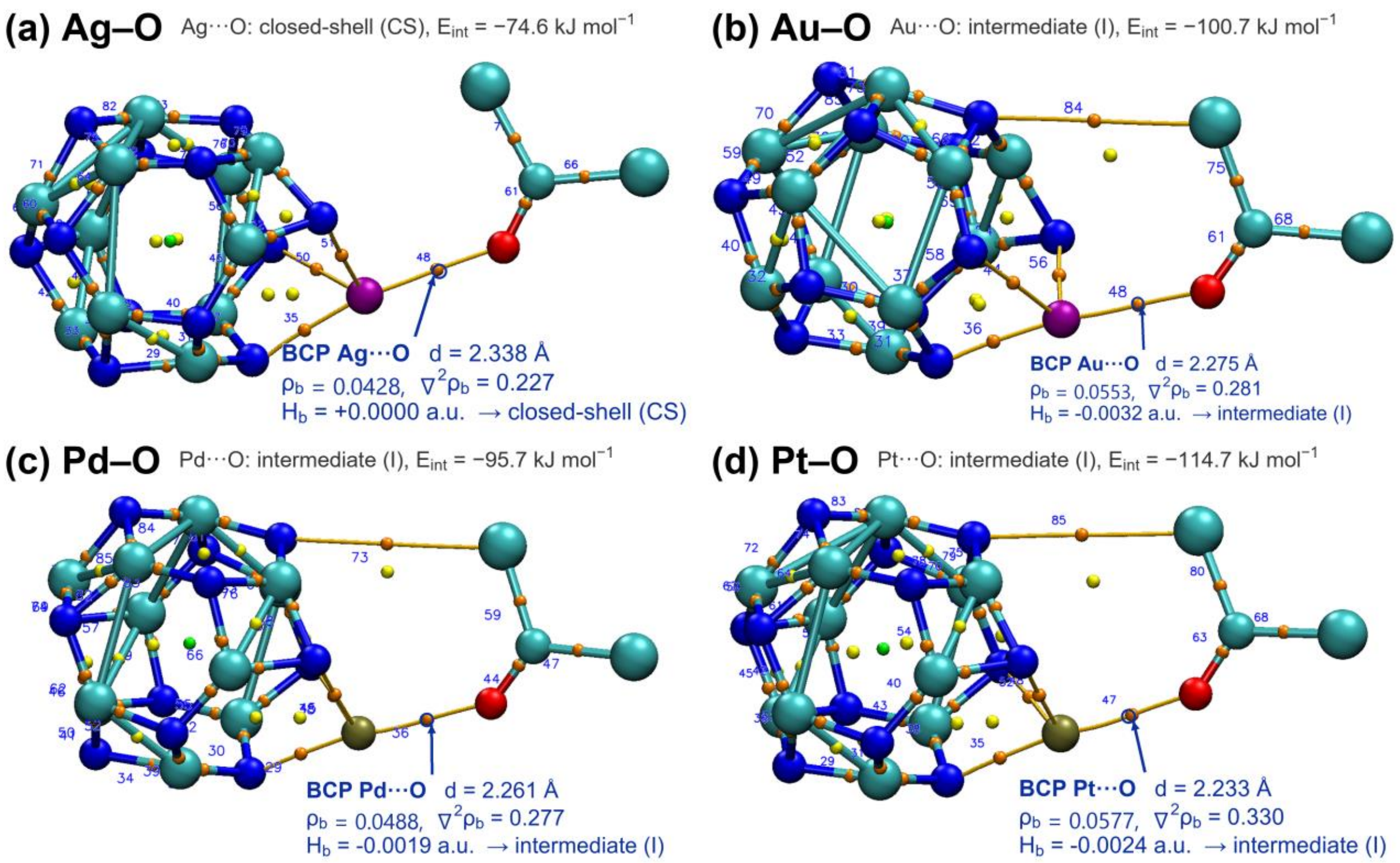

**Figure 6.** QTAIM molecular graphs of the (a) Ag–O, (b) Au–O, (c) Pd–O and (d) Pt–O complexes. Orange points are bond critical points, yellow points ring critical points and green points cage critical points; blue numbers label the atoms and bond paths as in Table S2. In these Multiwfn renderings N is blue, B, C and Cl are cyan, O is red, Ag and Au are magenta and Pd and Pt are olive. The cage···adsorbate bond critical point of each complex is circled and annotated with its distance, electron density ($\rho_b$), Laplacian ($\nabla^2\rho_b$), total energy density ($H_b$) and interaction type; the panel header gives the Espinosa interaction energy $E_{int} = V_b/2$.

Non-covalent interaction (NCI-RDG) analysis was employed [53] to visualise and quantify weak interactions via the reduced density gradient:

$$s = \frac{|\nabla\rho|}{2(3\pi^2)^{1/3}\,\rho^{4/3}} \tag{20}$$

Regions of low s mark non-covalent interactions. The sign of the second eigenvalue of the electron density Hessian ($\lambda_2$) separates attractive from repulsive contacts. The NCI-RDG scatter plots and isosurfaces of the oxygen-side complexes (Figure 7) and of the chlorine-side complexes (Figure S4) show two distinct features between the cage and $COCl_2$, both labelled in the figures. The first is a small blue disc at the metal···O or metal···Cl contact, coinciding with the bond critical point of Table S2; its spike in the scatter plot lies at sign($\lambda_2$)ρ ≈ −0.03 to −0.06 a.u., at the attractive edge of the plotted range, and is the NCI signature of the closed-shell and intermediate contacts identified by QTAIM. The second is a broad green disc between the cage surface and the phosgene backbone, whose spike at $\text{sign}(\lambda_2)\rho \approx -0.005$ to $-0.01$ a.u. is characteristic of van der Waals attraction. Au–Cl shows only the green N···Cl disc, consistent with the absence of an Au···Cl bond path, and Ag–Cl a single blue-green disc for its weak Ag···Cl contact. Apart from these localised contact spikes, no strongly negative $\text{sign}(\lambda_2)\rho$ values appear, so there is no covalent bonding. This agrees with the closed-shell character identified by AIM. The interaction between the doped nanocage and phosgene is therefore predominantly electrostatic and van der Waals in character, with the metal···O (or metal···Cl) contact supplying the directional component.

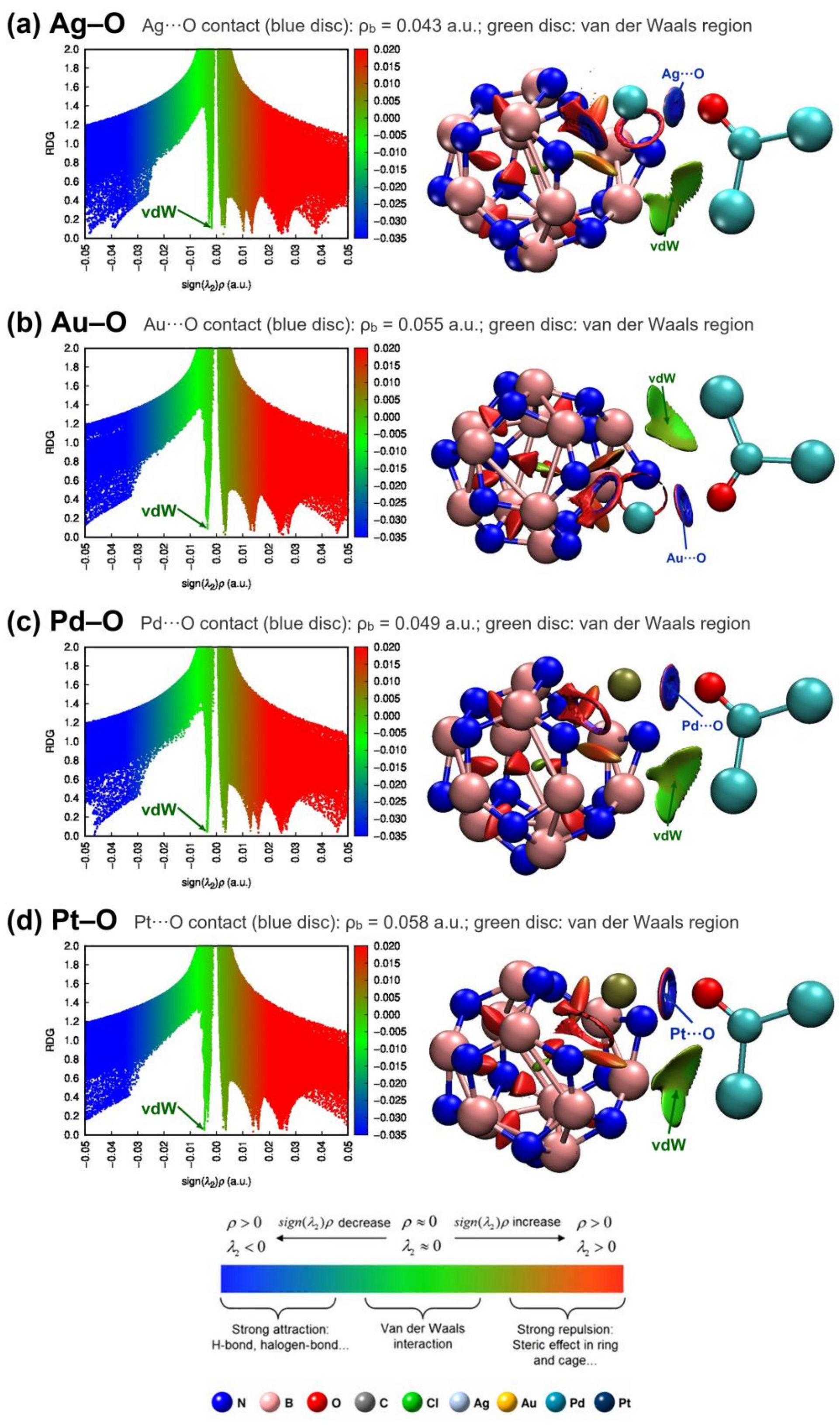

**Figure 7.** Quantum theory of atoms in molecules (QTAIM) and reduced density gradient (RDG) analysis of the (a) $AgB_{11}N_{12}$–$COCl_2$, (b) $AuB_{11}N_{12}$–$COCl_2$, (c) $PdB_{11}N_{12}$–$COCl_2$ and (d) $PtB_{11}N_{12}$–$COCl_2$ complexes with phosgene adsorbed near the oxygen side: identification of critical points and characterization of non-covalent interactions. The colour scale gives the interpretation of sign($\lambda_2$)·ρ. Blue labels mark the cage⋯adsorbate contact that carries a QTAIM bond path (Table S2) and green labels the van der Waals region; the corresponding spike is marked in each scatter plot.

### 3.14 Thermochemistry of Adsorption

Table S3 collects the vibrational analysis. All sixteen structures — the pristine and four doped cages, free phosgene, and the ten adsorption complexes — return exclusively real frequencies, confirming each as a genuine minimum on the potential energy surface rather than a saddle point. The bare cages are stiff, their lowest modes lying between 55.8 and 324.7 $cm^{-1}$, whereas every complex acquires a set of very soft modes, down to 2.9 $cm^{-1}$ in B–Cl and 6.2 $cm^{-1}$ in Au–Cl. These are the hindered translations and rotations of the adsorbate against the cage, and their softness is itself a signature of physisorption, since a chemisorbed molecule would be held by far stiffer modes. For the open-shell palladium and platinum systems the spin contamination is negligible: $\langle S^2 \rangle$ lies between 0.7522 and 0.7717 before annihilation and returns to 0.7500 or 0.7501 after it, against the exact doublet value of 0.75, so the unrestricted description of these systems is sound.

Table 8 converts these frequencies into the thermochemistry of adsorption at 298.15 K. Zero-point energy weakens the binding by 1.1–4.7 kJ $mol^{-1}$ and the thermal correction to the enthalpy by a further 1.5–4.1 kJ $mol^{-1}$, so ΔH is consistently smaller in magnitude than ΔE, running from −7.68 kJ $mol^{-1}$ for B–Cl to −62.91 kJ $mol^{-1}$ for Pt–O. The entropy term is much larger and it dominates. Adsorption removes the three translational and three rotational degrees of freedom of gaseous phosgene, so $\Delta S_{ads}$ is negative throughout, from −58.6 J $mol^{-1}$ $K^{-1}$ for the loosely bound B–Cl to −161.1 J $mol^{-1}$ $K^{-1}$ for Au–O. At 298.15 K this costs between 17.5 and 48.0 kJ $mol^{-1}$, comparable to or larger than the entire enthalpy of binding.

The soft modes that make adsorption entropically expensive are also the ones for which the harmonic approximation fails, so Table 8 reports the free energy both in the rigid-rotor harmonic-oscillator approximation and with Grimme's quasi-harmonic correction. The correction is not small, and it acts consistently in one direction: it raises ΔG by between 6.83 kJ $mol^{-1}$ (Pt–Cl) and 18.10 kJ $mol^{-1}$ (B–Cl). It does so because a free rotor carries less entropy than a harmonic oscillator of vanishing frequency, so removing the spurious low-frequency entropy of the complex makes $\Delta S_{ads}$ more negative and adsorption less favourable. The quasi-harmonic values are the more trustworthy of the two and are used throughout what follows; the harmonic values are retained in Table 8 so that the size of the correction remains visible.

The result reshapes the picture given by the electronic energies alone. On ΔE the pristine cage appears to bind phosgene weakly but appreciably, at −24.90 kJ $mol^{-1}$ for B–O. On $\Delta G_{qh}$ it does not bind it at all: B–O and B–Cl are strongly non-spontaneous, at +23.78 and +27.89 kJ $mol^{-1}$. Doping is therefore not a refinement of an already functioning material but a precondition for any binding whatever. Among the doped systems only Pt–O ($\Delta G_{qh}$ = −8.84 kJ $mol^{-1}$) and Pd–O (−7.23 kJ $mol^{-1}$) are spontaneous at room temperature. Four more — Au–O (+0.15), Pt–Cl (+0.24), Ag–O (+0.63) and Pd–Cl (+2.03 kJ $mol^{-1}$) — lie within about 2 kJ $mol^{-1}$ of equilibrium, which is well inside the uncertainty of the method and means only that they are marginal. Ag–Cl (+13.24) and Au–Cl (+30.21 kJ $mol^{-1}$) do not bind. For the oxygen-facing series the ordering by free energy, Pt > Pd > Au > Ag, is identical to the ordering by counterpoise-corrected electronic energy, so the conclusion that platinum is the best dopant survives the addition of entropy. What changes is that platinum's margin over the other metals widens, and that the chlorine-facing configurations drop out as viable binding modes.

**Table 8.** Thermochemistry of $COCl_2$ adsorption at 298.15 K and 1 atm. $\Delta G_{RRHO}$ uses the rigid-rotor harmonic-oscillator approximation; $\Delta G_{qh}$ applies Grimme's quasi-harmonic treatment to the vibrational entropy, with modes below about 100 $cm^{-1}$ interpolated towards free rotors. Negative values indicate spontaneous adsorption.

| Configuration | $\Delta E$ (kJ $mol^{-1}$) | $\Delta E$ + ZPE (kJ $mol^{-1}$) | $\Delta H$ (kJ $mol^{-1}$) | $\Delta S$ (J $mol^{-1}$ $K^{-1}$) | $-T\Delta S$ (kJ $mol^{-1}$) | $\Delta G_{RRHO}$ (kJ $mol^{-1}$) | $\Delta G_{qh}$ (kJ $mol^{-1}$) |
|---|---|---|---|---|---|---|---|
| B–O | −24.90 | −22.86 | −19.87 | −114.7 | 34.20 | 14.33 | 23.78 |
| B–Cl | −12.82 | −11.75 | −7.68 | −58.6 | 17.48 | 9.80 | 27.89 |
| Ag–O | −57.83 | −54.16 | −51.98 | −147.7 | 44.03 | −7.95 | 0.63 |
| Ag–Cl | −43.38 | −40.99 | −38.07 | −138.7 | 41.35 | 3.28 | 13.24 |
| Au–O | −61.06 | −56.34 | −54.85 | −161.1 | 48.02 | −6.84 | 0.15 |
| Au–Cl | −19.49 | −18.25 | −14.33 | −94.0 | 28.04 | 13.71 | 30.21 |
| Pd–O | −67.19 | −62.85 | −61.06 | −154.9 | 46.18 | −14.88 | −7.23 |
| Pd–Cl | −56.21 | −53.41 | −50.90 | −151.7 | 45.24 | −5.66 | 2.03 |
| Pt–O | −69.09 | −64.80 | −62.91 | −154.2 | 45.98 | −16.93 | −8.84 |
| Pt–Cl | −60.87 | −56.59 | −54.54 | −160.8 | 47.95 | −6.59 | 0.24 |

## 3.15 Sensor Performance and Operating Window

Table 9 makes the operating window explicit. The equilibrium constant for adsorption at 298.15 K spans nearly seven orders of magnitude across the ten configurations, from $5.09 \times 10^{-6}$ for Au–Cl to 35.4 for Pt–O. Only Pt–O and Pd–O exceed unity, and their coverages under one atmosphere of phosgene, 0.972 and 0.949, show that they would saturate in a pure analyte stream. Equivalently, half-coverage is reached at a partial pressure of 0.028 atm for Pt–O and 0.054 atm for Pd–O, against $1.46 \times 10^4$ atm for B–O — a pressure with no physical realisation, which is another way of saying that the pristine cage never binds.

The desorption temperature completes the picture. Pt–O holds phosgene up to 342.7 K (69.6 °C) and Pd–O up to 334.9 K (61.8 °C), so both operate at ambient temperature and regenerate on mild heating, which is exactly the behaviour a reusable sensor requires. The remaining doped configurations have $T_{des}$ within a few degrees of room temperature — Au–O at 24.2 °C, Pt–Cl at 23.8 °C, Ag–O at 21.7 °C and Pd–Cl at 14.6 °C — so they sit on the adsorption–desorption boundary and would respond only weakly and irreproducibly. Ag–Cl, Au–Cl and both pristine configurations desorb far below ambient and are not viable at all.

These thermodynamic $T_{des}$ values complement the Arrhenius recovery times of Table 6. The two answer different questions: $\tau$ shows that, once bound, phosgene leaves every cage within milliseconds at room temperature, whereas $T_{des}$ decides whether the cage retains it at all. Together they describe a material that binds phosgene reversibly at room temperature and gives it up rapidly on gentle warming.

**Table 9.** Thermodynamic sensing metrics derived from the quasi-harmonic free energies at 298.15 K. K is the equilibrium constant for adsorption, $p(\theta = ½)$ the phosgene partial pressure giving half-monolayer Langmuir coverage, $\theta$ the coverage under one atmosphere of phosgene, and $T_{des}$ the temperature at which $\Delta G_{ads}$ changes sign.

| Configuration | $\Delta G_{qh}$ (kJ $mol^{-1}$) | $K$ (298.15 K) | $p(\theta = ½)$ (atm) | θ at 1 atm | $T_{des}$ (K) | $T_{des}$ (°C) |
|---|---|---|---|---|---|---|
| B–O | 23.78 | $6.84 \times 10^{-5}$ | $1.46 \times 10^{4}$ | 0.000 | 154.0 | −119.1 |
| B–Cl | 27.89 | $1.30 \times 10^{-5}$ | $7.70 \times 10^{4}$ | 0.000 | 98.8 | −174.3 |
| Ag–O | 0.63 | $7.74 \times 10^{-1}$ | $1.29 \times 10^{0}$ | 0.436 | 294.9 | 21.7 |
| Ag–Cl | 13.24 | $4.78 \times 10^{-3}$ | $2.09 \times 10^{2}$ | 0.005 | 229.0 | −44.1 |
| Au–O | 0.15 | $9.41 \times 10^{-1}$ | $1.06 \times 10^{0}$ | 0.485 | 297.4 | 24.2 |
| Au–Cl | 30.21 | $5.09 \times 10^{-6}$ | $1.96 \times 10^{5}$ | 0.000 | 120.6 | −152.5 |
| Pd–O | −7.23 | $1.85 \times 10^{1}$ | $5.41 \times 10^{-2}$ | 0.949 | 334.9 | 61.8 |
| Pd–Cl | 2.03 | $4.42 \times 10^{-1}$ | $2.26 \times 10^{0}$ | 0.306 | 287.7 | 14.6 |

| Pt–O | −8.84 | $3.54 \times 10^{1}$ | $2.83 \times 10^{-2}$ | 0.972 | 342.7 | 69.6 |
|---|---|---|---|---|---|---|
| Pt–Cl | 0.24 | $9.07 \times 10^{-1}$ | $1.10 \times 10^{0}$ | 0.476 | 296.9 | 23.8 |

One consequence of this weak binding deserves to be stated plainly. At the 0.1 ppm occupational exposure limit for phosgene, the equilibrium coverage on Pt–O is of order $10^{-6}$, so the device operates deep in the Henry regime rather than anywhere near saturation. This is not a defect. In that regime the coverage, and with it the conductivity change, is linear in analyte concentration, which is precisely the response a quantitative sensor requires; saturation would instead destroy both the dynamic range and the reversibility. The trade-off is one of absolute sensitivity, which has to be recovered in the transduction electronics rather than by stronger chemisorption.

Bringing the thermodynamic and electronic criteria together, Pt–O is the only configuration that satisfies both. It is the strongest adsorber in the series ($\Delta E = -69.09$ kJ mol$^{-1}$, or $-60.71$ kJ mol$^{-1}$ after counterpoise correction); it is one of only two configurations with a negative free energy of adsorption at room temperature ($\Delta G_{qh} = -8.84$ kJ mol$^{-1}$, $K = 35.4$); it regenerates at a practical 69.6 °C; and it modulates the energy gap by +18.42%. Pd–O is the closest alternative on every count. The silver systems remain the most electrophilic ($\omega = 27.24$ eV for Ag–Cl) and give the narrowest post-adsorption gaps (1.33 eV for Ag–Cl and 1.37 eV for Ag–O), which would make them the better resistive transducers were they able to hold the analyte; with $\Delta G_{qh}$ of +13.24 and +0.63 kJ mol$^{-1}$ they cannot, so that electronic advantage is not realisable. Au–O gives the largest work function change (−9.45%) but sits on the adsorption threshold. On balance, platinum doping is the effective strategy for $COCl_2$ detection with $B_{12}N_{12}$ nanocages, and the free-energy analysis strengthens rather than merely preserves that conclusion, while removing silver from contention.

## 4. Conclusion

Boron-site substitution is the optimal strategy for noble-metal doping of $B_{12}N_{12}$ nanocages. It delivers substantially larger HOMO–LUMO gap reductions (69–82%) than nitrogen-site doping (41–67%) and a far stronger electrophilic character. Doping with Ag, Au, Pd, and Pt sharply enhances global reactivity, narrowing the pristine gap from 6.74 eV to 1.19–2.10 eV and raising the electrophilicity index from 3.14 eV to 13.27–31.79 eV.

Multiple adsorption starting geometries were tested for each system and all converged to the X–O and X–Cl configurations reported, confirming these as the global minima.

Pristine $B_{12}N_{12}$ interacts with $COCl_2$ only weakly on an electronic-energy basis ($\Delta E = -12.8$ to $-24.9$ kJ mol$^{-1}$), and once entropy is included it does not bind phosgene at all: its free energy of adsorption at 298 K is +23.8 kJ mol$^{-1}$ for B–O and +27.9 kJ mol$^{-1}$ for B–Cl. Each of the four noble metal-doped nanocages binds it more strongly in its preferred orientation, the eight doped configurations clustering between −19.5 kJ mol$^{-1}$ (Au–Cl) and −69.1 kJ mol$^{-1}$ (Pt–O), or −14.0 to −60.7 kJ mol$^{-1}$ after counterpoise correction. Dispersion supplies roughly 40% to 70% of that binding. Natural population analysis puts the charge transfer at 0.33 e or less, and QTAIM classifies every cage···adsorbate bond critical point as closed-shell or intermediate, so the interaction is physisorption in every case. Vibrational analysis confirms all sixteen structures as true minima and puts the entropy penalty of adsorption at 59–161 J mol$^{-1}$ K$^{-1}$, which costs 17–48 kJ mol$^{-1}$ at room temperature and is the decisive term: on a quasi-harmonic free-energy basis only Pt–O (−8.8 kJ mol$^{-1}$) and Pd–O (−7.2 kJ mol$^{-1}$) adsorb phosgene spontaneously at 298 K. This coherence contrasts with the wide dispersion seen across the full 3d transition metal series. The noble metals occupy a moderate adsorption window: strong enough for reliable phosgene detection, yet weak enough to avoid the irreversible over-binding that blocks sensor regeneration in early transition metals.

Continuum solvation confirms that every cage and every complex is stabilised in water, with platinum the most strongly solvated. Aqueous binding is weaker than gas-phase binding throughout, and Pt–Cl is marginally unbound under a continuum model applied to its fixed gas-phase geometry, which should be treated as a limitation of that approximation rather than as a prediction that the complex cannot form. The free energies reported here are gas-phase quantities, since no frequency analysis was carried out in solution; the aqueous results should therefore be read as electronic-energy trends rather than as solution thermodynamics.

Platinum is the strongest adsorber under every criterion applied — electronic energy, counterpoise-corrected energy, enthalpy and free energy — and $PtB_{11}N_{12}$ retains the most structural stability of the four doped cages. The Pt–O complex is spontaneous at room temperature ($\Delta G_{qh} = -8.8$ kJ mol$^{-1}$, K = 35.4), desorbs at a practical 69.6 °C, and modulates the energy gap by +18.4%, so it combines binding, reversibility and a measurable signal in one configuration. Pd–O is the only close alternative. $AgB_{11}N_{12}$ gives the narrowest post-adsorption gap (1.33 eV) and the highest electrophilicity ($\omega$ = 27.24 eV), and would on those grounds be the better resistive transducer, but free energies of adsorption of +13.2 and +0.6 kJ mol$^{-1}$ show that it cannot retain phosgene at room temperature, so that advantage cannot be realised. $AuB_{11}N_{12}$ gives the largest work function variation (−9.45% for Au–O) but sits on the adsorption threshold. Boron-site platinum doping is therefore the effective route to phosgene detection with $B_{12}N_{12}$ nanocages. It provides a firm theoretical basis for the rational design of BN-based gas sensors, and also a caution: for weakly physisorbed analytes, adsorption energies alone can substantially overstate how many candidate materials will work in practice.

## CRediT authorship contribution statement

Shahariar Chowdhury: Conceptualization, Methodology, Software, Investigation, Formal analysis, Data curation, Visualization, Writing – original draft. Mohammad Abdul Matin: Supervision, Writing – review & editing. Samiran Bhattacharjee: Supervision, Writing – review & editing. Ishtiaque M. Syed: Supervision, Resources, Project administration, Writing – review & editing.

## Acknowledgements

The authors acknowledge the computational resources provided by the Centre for Advanced Research in Sciences (CARS), University of Dhaka, Dhaka, Bangladesh.

## Conflict of Interest

The authors declare no conflict of interest.

**Supplementary Information**

# Adsorption of Phosgene Gas on Pristine and Noble Metal-Doped $B_{12}N_{12}$ Nanocages: Insights from Density Functional Theory

Shahariar Chowdhury[1,2], Mohammad Abdul Matin[2], Samiran Bhattacharjee[2*] and Ishtiaque M. Syed[1*]

[1]*Centre for Advanced Research in Sciences (CARS), University of Dhaka, Dhaka, Bangladesh*
[2]*Materials Physics Laboratory, Department of Physics, University of Dhaka, Dhaka, Bangladesh*

*Corresponding authors' e-mail: imsyed@du.ac.bd (IMS), s.bhattacharjee@du.ac.bd (SB)

## Contents

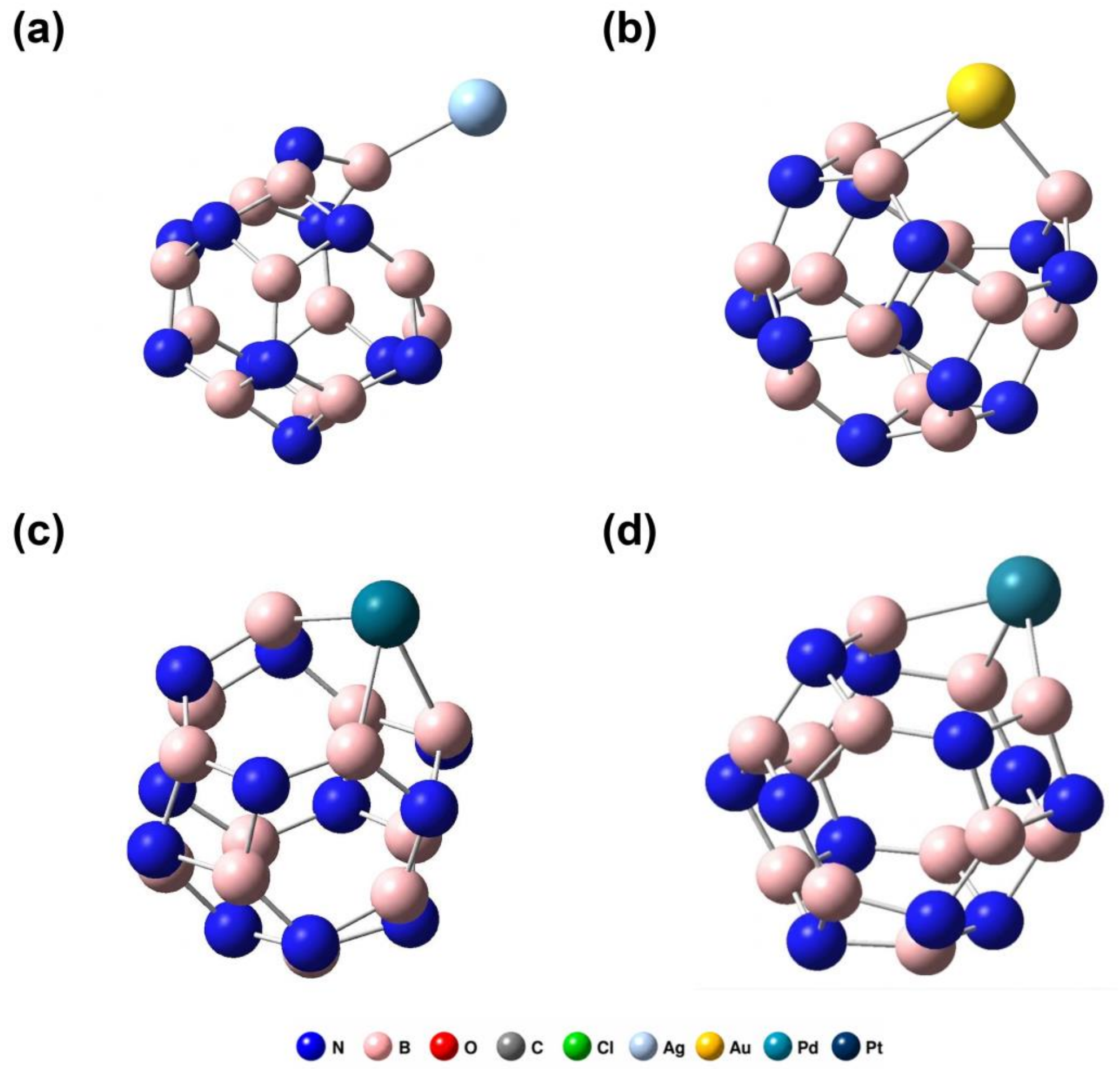

**Figure S1.** Optimised structural configurations of the nitrogen-site doped (a) $AgN_{11}B_{12}$, (b) $AuN_{11}B_{12}$, (c) $PdN_{11}B_{12}$ and (d) $PtN_{11}B_{12}$ nanoclusters.

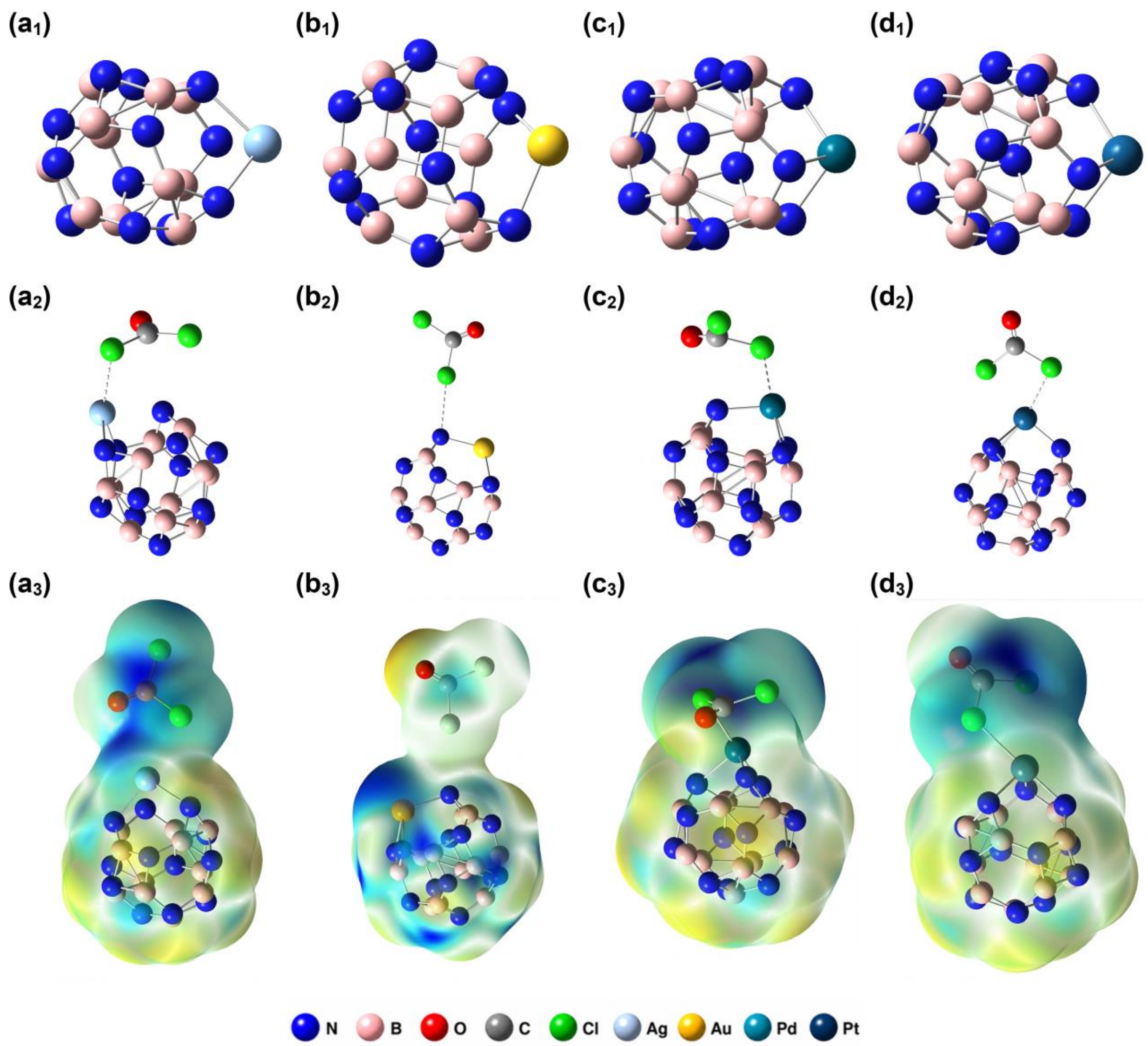

**Figure S2.** Representations of the optimised geometry of $AgB_{11}N_{12}$ ($a_1$), $AuB_{11}N_{12}$ ($b_1$), $PdB_{11}N_{12}$ ($c_1$) and $PtB_{11}N_{12}$ ($d_1$); optimised geometry of $AgB_{11}N_{12}$ ($a_2$), $AuB_{11}N_{12}$ ($b_2$), $PdB_{11}N_{12}$ ($c_2$) and $PtB_{11}N_{12}$ ($d_2$) with phosgene gas ($COCl_2$) adsorbed near the chlorine side; and MEP of $AgB_{11}N_{12}$ ($a_3$), $AuB_{11}N_{12}$ ($b_3$), $PdB_{11}N_{12}$ ($c_3$) and $PtB_{11}N_{12}$ ($d_3$) with phosgene gas ($COCl_2$) adsorbed near the chlorine side.

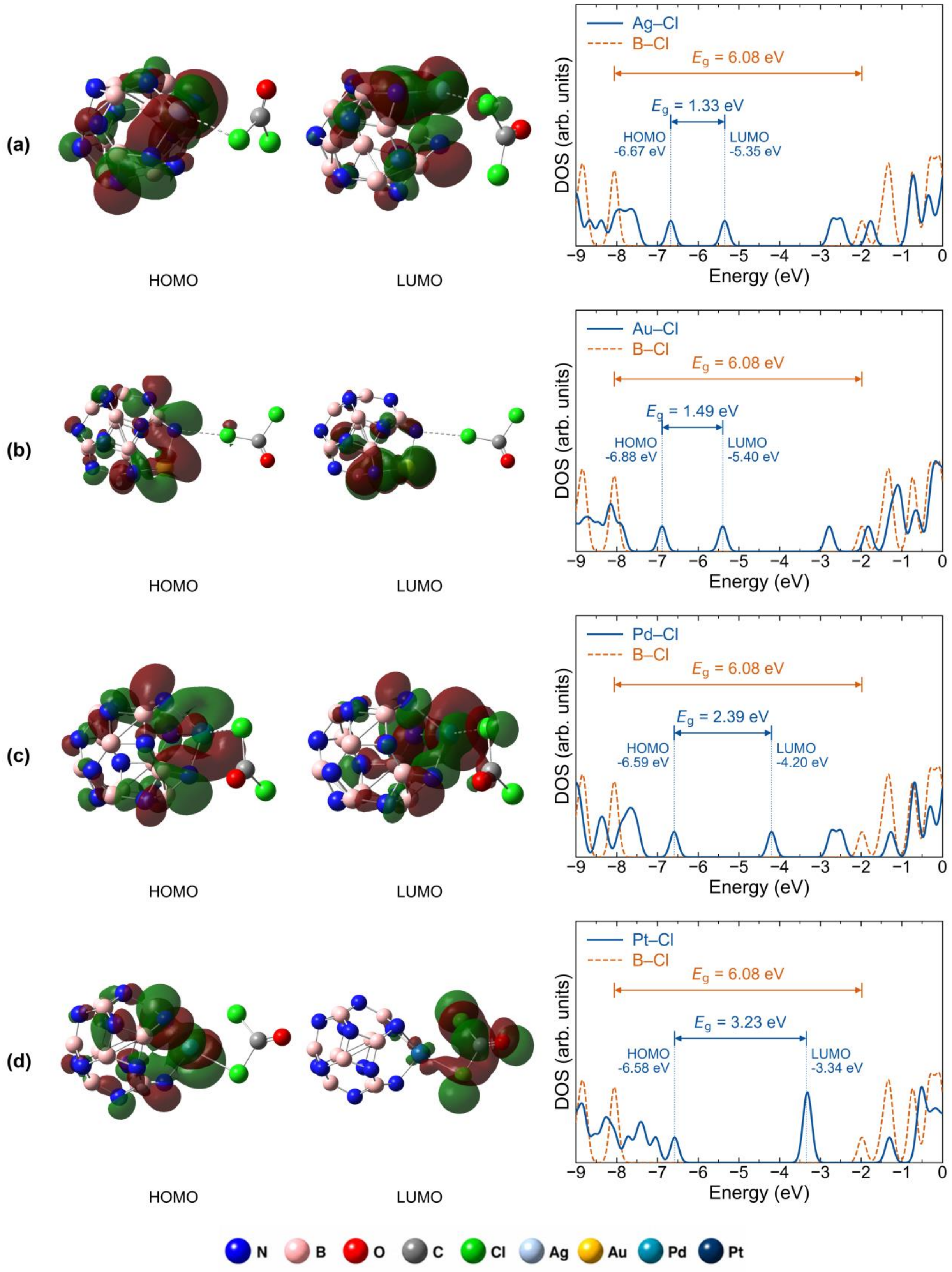

**Figure S3.** Representations of the HOMO, LUMO and DOS of the optimised geometry of $AgB_{11}N_{12}$ (a), $AuB_{11}N_{12}$ (b), $PdB_{11}N_{12}$ (c) and $PtB_{11}N_{12}$ (d) with phosgene gas ($COCl_2$) adsorbed near the chlorine side. The dashed orange trace in each DOS panel is the pristine B–Cl complex.

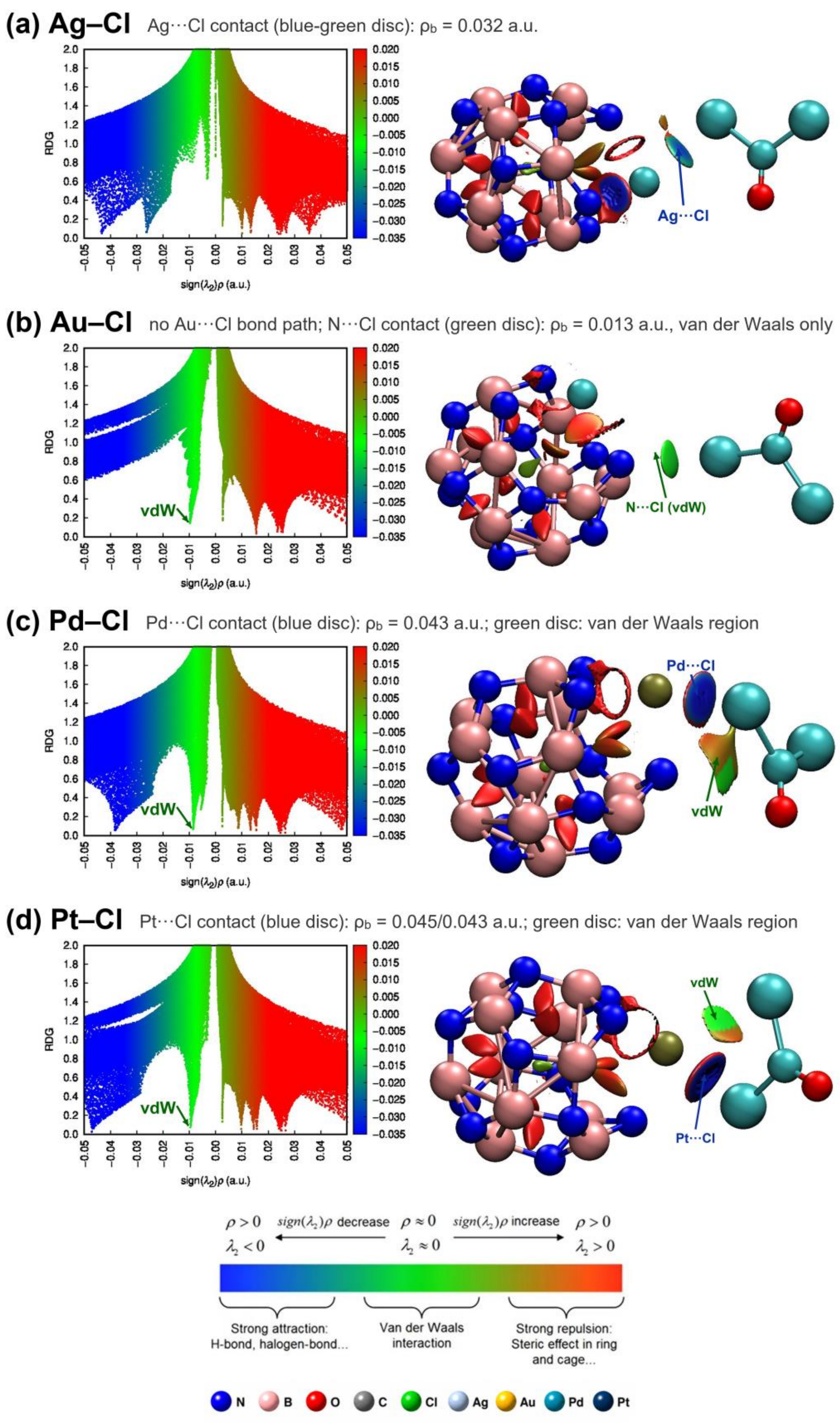

**Figure S4.** Quantum theory of atoms in molecules (QTAIM) and reduced density gradient (RDG) analysis of the (a) $AgB_{11}N_{12}$–$COCl_2$, (b) $AuB_{11}N_{12}$–$COCl_2$, (c) $PdB_{11}N_{12}$–$COCl_2$ and (d) $PtB_{11}N_{12}$–$COCl_2$ complexes with phosgene adsorbed near the chlorine side: identification of critical points and characterisation of non-covalent interactions. Blue labels mark the cage···adsorbate contact that carries a QTAIM bond path (Table S2) and green labels the van der Waals region; the corresponding spike is marked in each scatter plot.

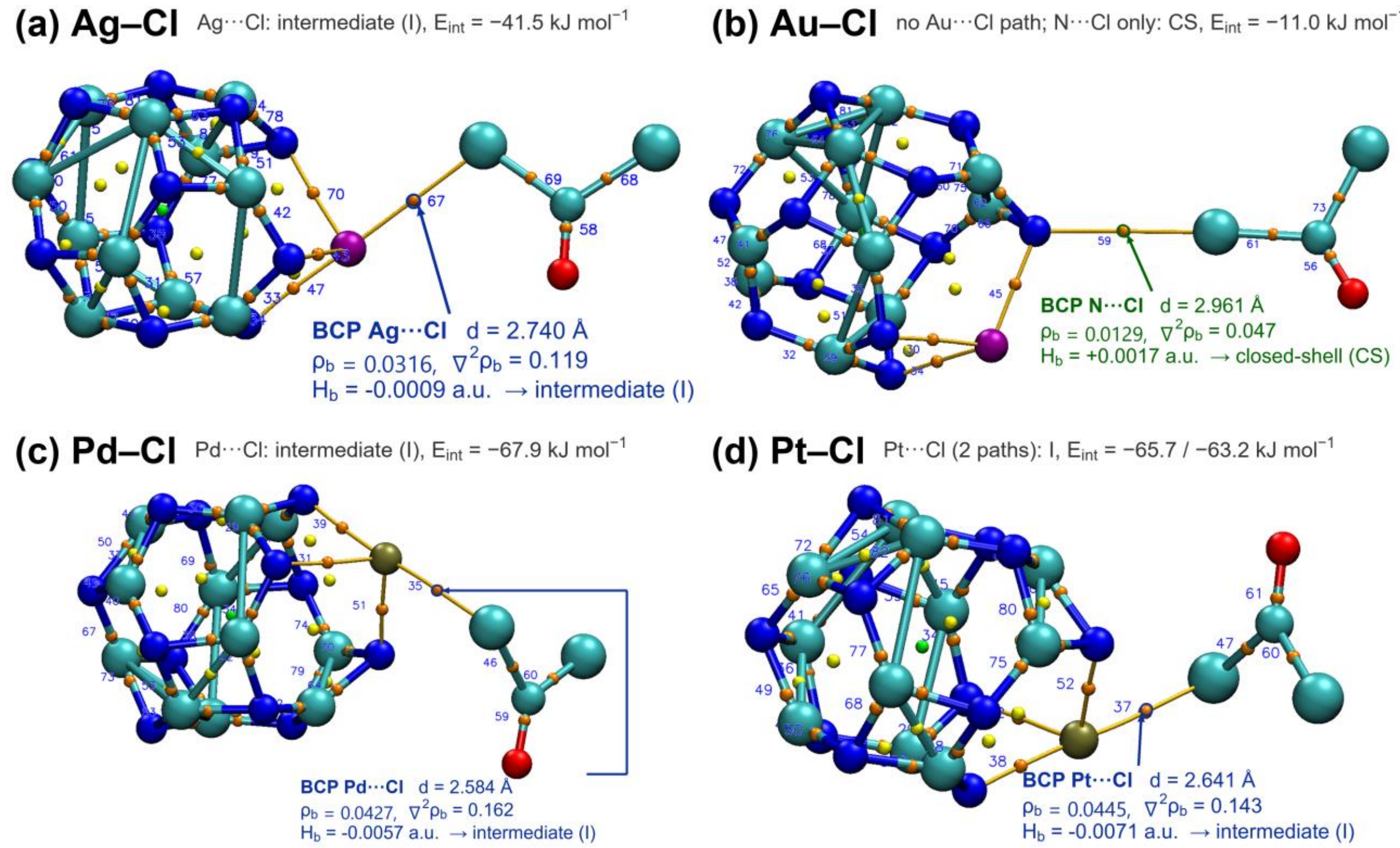

**Figure S5.** QTAIM molecular graphs of the (a) Ag–Cl, (b) Au–Cl, (c) Pd–Cl and (d) Pt–Cl complexes. Orange points are bond critical points, yellow points ring critical points and green points cage critical points; blue numbers label the atoms and bond paths as in Table S2. N is blue, B, C and Cl are cyan, O is red, Ag and Au are magenta and Pd and Pt are olive. The cage···adsorbate bond critical point is circled and annotated with its distance, $\rho_b$, $\nabla^2\rho_b$, $H_b$ and interaction type, and the panel header gives the Espinosa interaction energy $E_{int} = V_b/2$; Au–Cl has no Au···Cl bond path and is linked to the cage only through an N···Cl contact.

## Supplementary Tables

**Table S1.** Total energies, solvation energies, adsorption energies and aqueous-phase recovery times of the nanocages and their $COCl_2$ complexes.

| System | *E*gas (Ha) | *E*water (Ha) | *E*sol (kJ mol−1) | *E*ads gas (kJ mol−1) | *E*ads aq (kJ mol−1) | *τ* aq (ms) |
|---|---|---|---|---|---|---|
| ***Free adsorbate*** | | | | | | |
| $COCl_2$ | −1033.81754 | −1033.82000 | −6.45 | -- | -- | -- |
| ***Undoped*** | | | | | | |
| $B_{12}N_{12}$ | −956.44764 | −956.45434 | −17.58 | -- | -- | -- |
| B–O | −1990.27467 | −1990.28326 | −22.56 | −24.9 | −23.4 | 1.27 × 10−5 |
| B–Cl | −1990.27006 | −1990.27877 | −22.86 | −12.8 | −11.6 | 1.10 × 10−7 |
| ***Ag-doped*** | | | | | | |
| $AgB_{11}N_{12}$ | −1077.18173 | −1077.20887 | −71.28 | -- | -- | -- |
| Ag–O | −2111.02129 | −2111.04191 | −54.14 | −57.8 | −34.2 | 9.93 × 10−4 |
| Ag–Cl | −2111.01579 | −2111.03780 | −57.78 | −43.4 | −23.4 | 1.27 × 10−5 |
| ***Au-doped*** | | | | | | |
| $AuB_{11}N_{12}$ | −1066.86882 | −1066.88803 | −50.45 | -- | -- | -- |
| Au–O | −2100.70962 | −2100.72766 | −47.37 | −61.1 | −51.5 | 1.07 × 100 |
| Au–Cl | −2100.69378 | −2100.71353 | −51.85 | −19.5 | −14.4 | 3.39 × 10−7 |
| ***Pd-doped*** | | | | | | |
| $PdB_{11}N_{12}$ | −1058.20691 | −1058.24181 | −91.64 | -- | -- | -- |
| Pd–O | −2092.05004 | −2092.07657 | −69.65 | −67.2 | −38.7 | 6.13 × 10−3 |
| Pd–Cl | −2092.04586 | −2092.07146 | −67.22 | −56.2 | −25.3 | 2.75 × 10−5 |
| ***Pt-doped*** | | | | | | |
| $PtB_{11}N_{12}$ | −1050.62359 | −1050.67026 | −122.51 | -- | -- | -- |
| Pt–O | −2084.46745 | −2084.50211 | −91.01 | −69.1 | −31.1 | 2.85 × 10−4 |
| Pt–Cl | −2084.46432 | −2084.48886 | −64.42 | −60.9 | 3.7 | n/a (unbound) |

**Table S2.** QTAIM topological parameters at the bond critical points of all doped adsorption configurations, computed from the B3LYP-D3(BJ) wavefunctions.

| Bond critical point | *d* (Å) | *ρ*b (a.u.) | ∇2*ρ*b (a.u.) | *G*b (a.u.) | *V*b (a.u.) | *H*b (a.u.) | \|*V*b\|/*G*b | Type; *E*int (kJ mol−1) |
|---|---|---|---|---|---|---|---|---|
| ***Ag–O (O-down)*** | | | | | | | | |
| N14–Ag28 | 2.082 | 0.0974 | 0.2570 | 0.0925 | −0.1207 | −0.02823 | 1.31 | I |
| N15–Ag28 | 2.207 | 0.0732 | 0.2678 | 0.0816 | −0.0963 | −0.01465 | 1.18 | I |
| N12–Ag28 | 2.383 | 0.0501 | 0.1808 | 0.0502 | −0.0551 | −0.00497 | 1.10 | I |
| O25···Ag28 | 2.338 | 0.0428 | 0.2273 | 0.0568 | −0.0568 | 0.00000 | 1.00 | CS; −74.6 |
| $C_{24}$–O25 | 1.193 | 0.4241 | 0.0380 | 0.7347 | −1.4599 | −0.72521 | 1.99 | I |
| $C_{24}$–Cl27 | 1.731 | 0.2083 | −0.2751 | 0.0767 | −0.2221 | −0.14544 | 2.90 | S |
| $C_{24}$–Cl26 | 1.738 | 0.2053 | −0.2679 | 0.0747 | −0.2163 | −0.14164 | 2.90 | S |
| ***Ag–Cl (Cl-down)*** | | | | | | | | |
| N22–Ag28 | 2.102 | 0.0925 | 0.2497 | 0.0878 | −0.1133 | −0.02543 | 1.29 | I |
| N13–Ag28 | 2.182 | 0.0770 | 0.2832 | 0.0873 | −0.1038 | −0.01649 | 1.19 | I |
| N19–Ag28 | 2.359 | 0.0528 | 0.1893 | 0.0534 | −0.0595 | −0.00608 | 1.11 | I |
| Cl26···Ag28 | 2.740 | 0.0316 | 0.1191 | 0.0307 | −0.0316 | −0.00092 | 1.03 | I; −41.5 |
| N19···$C_{24}$ | 3.141 | 0.0076 | 0.0251 | 0.0052 | −0.0042 | 0.00104 | 0.80 | CS; −5.5 |
| $C_{24}$–O25 | 1.169 | 0.4494 | 0.2468 | 0.8505 | −1.6393 | −0.78882 | 1.93 | I |
| $C_{24}$–Cl27 | 1.734 | 0.2054 | −0.2645 | 0.0757 | −0.2174 | −0.14179 | 2.87 | S |
| $C_{24}$–Cl26 | 1.819 | 0.1729 | −0.1564 | 0.0621 | −0.1634 | −0.10123 | 2.63 | S |
| ***Au–O (O-down)*** | | | | | | | | |
| N12–Au28 | 2.017 | 0.1296 | 0.2676 | 0.1108 | −0.1547 | −0.04391 | 1.40 | I |
| N15–Au28 | 2.180 | 0.0905 | 0.2587 | 0.0886 | −0.1125 | −0.02392 | 1.27 | I |
| N14–Au28 | 2.333 | 0.0674 | 0.1852 | 0.0609 | −0.0756 | −0.01465 | 1.24 | I |
| O25···Au28 | 2.275 | 0.0553 | 0.2808 | 0.0735 | −0.0767 | −0.00325 | 1.04 | I; −100.7 |
| $C_{24}$–O25 | 1.198 | 0.4179 | 0.0225 | 0.7149 | −1.4242 | −0.70931 | 1.99 | I |
| $C_{24}$–Cl27 | 1.723 | 0.2118 | −0.2877 | 0.0782 | −0.2284 | −0.15018 | 2.92 | S |
| $C_{24}$–Cl26 | 1.732 | 0.2079 | −0.2780 | 0.0756 | −0.2207 | −0.14513 | 2.92 | S |
| ***Au–Cl (Cl-down)*** | | | | | | | | |
| N19–Au28 | 2.127 | 0.1020 | 0.2206 | 0.0861 | −0.1171 | −0.03095 | 1.36 | I |
| N22–Au28 | 2.130 | 0.1013 | 0.2203 | 0.0857 | −0.1164 | −0.03064 | 1.36 | I |
| N13–Au28 | 2.151 | 0.0960 | 0.3035 | 0.1021 | −0.1284 | −0.02625 | 1.26 | I |
| N13···Cl26 | 2.961 | 0.0129 | 0.0471 | 0.0101 | −0.0084 | 0.00171 | 0.83 | CS; −11.0 |
| $C_{24}$–O25 | 1.176 | 0.4421 | 0.1602 | 0.8111 | −1.5821 | −0.77101 | 1.95 | I |
| $C_{24}$–Cl26 | 1.756 | 0.1984 | −0.2380 | 0.0723 | −0.2041 | −0.13178 | 2.82 | S |
| $C_{24}$–Cl27 | 1.773 | 0.1901 | −0.2117 | 0.0694 | −0.1916 | −0.12227 | 2.76 | S |
| ***Pd–O (O-down)*** | | | | | | | | |
| Pd12–N20 | 1.991 | 0.1252 | 0.2926 | 0.1192 | −0.1652 | −0.04603 | 1.39 | I |
| Pd12–N14 | 2.033 | 0.1130 | 0.2868 | 0.1106 | −0.1495 | −0.03891 | 1.35 | I |
| Pd12–N13 | 2.284 | 0.0669 | 0.1900 | 0.0609 | −0.0743 | −0.01340 | 1.22 | I |
| Pd12···O26 | 2.261 | 0.0488 | 0.2765 | 0.0710 | −0.0729 | −0.00190 | 1.03 | I; −95.7 |
| C25–O26 | 1.194 | 0.4221 | 0.0419 | 0.7303 | −1.4501 | −0.71979 | 1.99 | I |
| C25–Cl28 | 1.728 | 0.2098 | −0.2804 | 0.0773 | −0.2248 | −0.14744 | 2.91 | S |
| C25–Cl27 | 1.737 | 0.2057 | −0.2704 | 0.0747 | −0.2170 | −0.14231 | 2.90 | S |
| ***Pd–Cl (Cl-down)*** | | | | | | | | |
| Pd12–N20 | 1.998 | 0.1233 | 0.2792 | 0.1153 | −0.1608 | −0.04551 | 1.39 | I |
| Pd12–N14 | 2.064 | 0.1060 | 0.2771 | 0.1047 | −0.1400 | −0.03538 | 1.34 | I |
| Pd12–N13 | 2.179 | 0.0836 | 0.2194 | 0.0778 | −0.1007 | −0.02294 | 1.29 | I |
| Pd12···Cl27 | 2.584 | 0.0427 | 0.1618 | 0.0461 | −0.0518 | −0.00566 | 1.12 | I; −67.9 |
| N14···C25 | 2.847 | 0.0129 | 0.0426 | 0.0093 | −0.0079 | 0.00137 | 0.85 | CS; −10.4 |
| C25–O26 | 1.166 | 0.4518 | 0.2732 | 0.8629 | −1.6575 | −0.79461 | 1.92 | I |
| C25–Cl28 | 1.733 | 0.2061 | −0.2674 | 0.0757 | −0.2183 | −0.14260 | 2.88 | S |
| C25–Cl27 | 1.839 | 0.1657 | −0.1353 | 0.0592 | −0.1522 | −0.09300 | 2.57 | S |
| ***Pt–O (O-down)*** | | | | | | | | |
| Pt12–N20 | 1.995 | 0.1369 | 0.3120 | 0.1268 | −0.1757 | −0.04884 | 1.39 | I |
| Pt12–N14 | 2.038 | 0.1240 | 0.2985 | 0.1159 | −0.1572 | −0.04128 | 1.36 | I |
| Pt12–N13 | 2.239 | 0.0840 | 0.2049 | 0.0732 | −0.0952 | −0.02200 | 1.30 | I |
| Pt12···O26 | 2.233 | 0.0577 | 0.3302 | 0.0850 | −0.0874 | −0.00241 | 1.03 | I; −114.7 |
| C25–O26 | 1.198 | 0.4172 | 0.0373 | 0.7164 | −1.4236 | −0.70712 | 1.99 | I |
| C25–Cl28 | 1.723 | 0.2117 | −0.2872 | 0.0783 | −0.2284 | −0.15008 | 2.92 | S |
| C25–Cl27 | 1.733 | 0.2076 | −0.2774 | 0.0753 | −0.2200 | −0.14469 | 2.92 | S |
| ***Pt–Cl (Cl-down)*** | | | | | | | | |
| Pt12–N14 | 1.977 | 0.1406 | 0.3373 | 0.1349 | −0.1854 | −0.05054 | 1.37 | I |
| Pt12–N20 | 2.014 | 0.1311 | 0.3007 | 0.1203 | −0.1654 | −0.04510 | 1.37 | I |
| Pt12–N13 | 2.545 | 0.0458 | 0.1225 | 0.0363 | −0.0420 | −0.00571 | 1.16 | I |
| Pt12···Cl27 | 2.641 | 0.0445 | 0.1429 | 0.0429 | −0.0500 | −0.00714 | 1.17 | I; −65.7 |
| Pt12···Cl28 | 2.657 | 0.0433 | 0.1375 | 0.0413 | −0.0482 | −0.00690 | 1.17 | I; −63.2 |
| C25–O26 | 1.165 | 0.4527 | 0.2853 | 0.8679 | −1.6645 | −0.79660 | 1.92 | I |
| C25–Cl28 | 1.785 | 0.1860 | −0.1946 | 0.0677 | −0.1840 | −0.11632 | 2.72 | S |
| C25–Cl27 | 1.791 | 0.1840 | −0.1881 | 0.0668 | −0.1806 | −0.11383 | 2.70 | S |

Interaction type follows Bader and Cremer–Kraka and is coded S, I or CS: S (shared, covalent) for $\nabla^2\rho_b < 0$ with $H_b < 0$; I (intermediate, partly covalent) for $\nabla^2\rho_b > 0$ with $H_b < 0$; and CS (closed-shell, electrostatic) for $\nabla^2\rho_b > 0$ with $H_b \geq 0$. Where a value follows the code it is $E_{int}$, the Espinosa estimate $V_b/2$, quoted only for the cage···adsorbate contacts, the only interactions for which that approximation applies. Every cage···adsorbate contact is closed-shell or intermediate and none is shared, consistent with the physisorption picture from the adsorption energies and the small NBO charge transfer of Table 4.

**Table S3.** Vibrational analysis and spin-state validation of all optimised structures. $N_{imag}$ is the number of imaginary frequencies; $\langle S^2 \rangle$ is quoted before and after spin annihilation for the open-shell doublets, against an exact value of 0.75.

| System | Normal modes | $N_{imag}$ | Lowest mode ($cm^{-1}$) | Highest mode ($cm^{-1}$) | ZPE (kJ $mol^{-1}$) | Multiplicity | $\langle S^2 \rangle$ before / after |
|---|---|---|---|---|---|---|---|
| $COCl_2$ | 6 | 0 | 303.4 | 1879.6 | 27.4 | 1 | -- |
| $B_{12}N_{12}$ | 66 | 0 | 324.7 | 1443.1 | 337.8 | 1 | -- |
| B–O | 78 | 0 | 16.0 | 1827.9 | 367.2 | 1 | -- |
| B–Cl | 78 | 0 | 2.9 | 1875.7 | 366.3 | 1 | -- |
| $AgB_{11}N_{12}$ | 66 | 0 | 77.8 | 1441.0 | 312.0 | 1 | -- |
| Ag–O | 78 | 0 | 13.3 | 1767.8 | 343.2 | 1 | -- |
| Ag–Cl | 78 | 0 | 15.9 | 1910.0 | 341.9 | 1 | -- |
| $AuB_{11}N_{12}$ | 66 | 0 | 55.8 | 1454.5 | 313.5 | 1 | -- |
| Au–O | 78 | 0 | 18.6 | 1735.4 | 345.6 | 1 | -- |
| Au–Cl | 78 | 0 | 6.2 | 1866.4 | 342.2 | 1 | -- |
| $PdB_{11}N_{12}$ | 66 | 0 | 95.5 | 1441.4 | 316.2 | 2 | 0.7533 / 0.7500 |
| Pd–O | 78 | 0 | 17.0 | 1762.2 | 348.0 | 2 | 0.7690 / 0.7501 |
| Pd–Cl | 78 | 0 | 16.6 | 1926.6 | 346.4 | 2 | 0.7717 / 0.7501 |
| $PtB_{11}N_{12}$ | 66 | 0 | 100.5 | 1444.9 | 317.0 | 2 | 0.7522 / 0.7500 |
| Pt–O | 78 | 0 | 20.0 | 1735.2 | 348.8 | 2 | 0.7577 / 0.7500 |
| Pt–Cl | 78 | 0 | 23.2 | 1935.1 | 348.7 | 2 | 0.7571 / 0.7500 |